\documentclass[aps,prb,superscriptaddress,showpacs,preprint]{revtex4}
\usepackage{amsmath,amsthm}
\usepackage{amsfonts}
\usepackage[utf8]{inputenc}
\usepackage{graphicx}
\usepackage{xcolor}
\usepackage{hyperref}
\usepackage{color}
\usepackage{bm}
\allowdisplaybreaks

\begin{document}

\title{Benchmarking the nonperturbative functional renormalization group approach on the random elastic manifold model in and out of equilibrium}

\author{Ivan Balog} \email{balog@ifs.hr}
\affiliation{Institute of Physics, P.O.Box 304, Bijeni\v{c}ka cesta 46, HR-10001 Zagreb, Croatia}

\author{Gilles Tarjus} \email{tarjus@lptmc.jussieu.fr}
\affiliation{LPTMC, CNRS-UMR 7600, Universit\'e Pierre et Marie Curie,
bo\^ite 121, 4 Pl. Jussieu, 75252 Paris c\'edex 05, France}

\author{Matthieu Tissier} \email{tissier@lptmc.jussieu.fr}
\affiliation{LPTMC, CNRS-UMR 7600, Universit\'e Pierre et Marie Curie,
bo\^ite 121, 4 Pl. Jussieu, 75252 Paris c\'edex 05, France}

\date{\today}

\begin{abstract}

Criticality in the class of disordered systems comprising the random-field Ising model (RFIM) and elastic manifolds in a random environment is controlled by zero-temperature fixed points that must be treated through a functional renormalization group. We apply the nonperturbative functional renormalization group approach that we have previously used to describe the RFIM in and out of equilibrium [Balog-Tarjus-Tissier, Phys. Rev. B {\bf 97}, 094204 (2018)] to the simpler and by now well-studied case of the random elastic manifold model. We recover the main known properties, critical exponents and scaling functions, of both the pinned phase of the manifold at equilibrium and the depinning threshold in the athermally and quasi-statically driven case for any dimension $0<d\leq 4$. This successful benchmarking of our theoretical approach gives strong support to the results that we have previously obtained for the RFIM, in particular concerning the distinct universality classes of the equilibrium and out-of-equilibrium (hysteresis) critical points below a critical dimension $d_{DR}\approx 5.1$.

\end{abstract}

\pacs{11.10.Hi, 75.40.Cx}

\maketitle

\section{Introduction}
\label{introduction}

There are classes of disordered models for which the long-distance physics is controlled, in a renormalization-group language, by zero-temperature fixed points, which entails that the disorder grows under renormalization and that the sample-to-sample fluctuations dominate the thermal fluctuations at large scale. Among others, this is the case of bulk systems in the presence of a random field and of elastic manifolds in a random environment.\cite{footnote_SG} Both types of models can be studied in equilibrium, with the (bulk) random-field systems showing a critical point corresponding to a paramagnetic to ferromagnetic transition above a lower critical dimension\cite{imry-ma75,nattermann98} and the random elastic manifold models being in a pinned phase where they are characterized by roughness on all scales.\cite{nattermann98,fisher86-pin} Both types of models can also be investigated at zero temperature when quasi-statically driven by an applied force or magnetic field. This leads to hysteresis and an out-of-equilibrium critical point for the random-field systems\cite{sethna93,sethna05} and to a depinning transition for the random elastic manifold models.\cite{fisher_dep,nattermann_dep,narayan92}

The combination of quenched disorder and zero-temperature fixed point leads to the presence of discontinuous, collective events, known as static avalanches or shocks in the equilibrium case and dynamic avalanches in the driven case. These events appear on all scales at criticality and manifest themselves in the correlation functions and the cumulants of the renormalized disorder as a singular dependence on the arguments that takes the form of linear cusps.\cite{fisher86-pin,nattermann_dep,narayan92,BBM,feldman,FRGchauve,FRGledoussal-chauve,tarjus04} When these cusps persist in the renormalized quantities up to the fixed point, they strongly influence the universal behavior and, in particular, they  lead to a breakdown of the dimensional-reduction results predicted by conventional perturbation theory and supersymmetry-based arguments for all of these models.  (Dimensional reduction here means that the critical behavior of the disordered model in dimension $d$ is the same as that of the corresponding pure model in dimension $d-2$.) Accounting for the effect of avalanches and of the ensuing cuspy behavior in a renormalization group (RG) setting requires a {\it functional} approach in which one does not merely follow the RG flow of a set of coupling constants but rather that of a set of functions.

In the case of the random elastic manifold models the upper critical dimension is $d_{uc}=4$ and it turns out that one can perform a {\it perturbative} functional RG treatment in $\epsilon=4-d$, both for the pinned phase at equilibrium and for the depinning transition.\cite{fisher86-pin,nattermann_dep,narayan92,FRGchauve,FRGledoussal-chauve} On the contrary in the random-field models the upper critical dimension in $d_{uc}=6$ but the transition from a regime where supersymmetry and dimensional reduction are valid to one where they break down due to avalanches and cusps takes place at a nontrivial critical dimension, {\it e.g.}, $d_{DR}\approx 5.1$ for the Ising version, which is only accessible through a {\it nonperturbative} functional RG.\cite{tarjus04,tissier06,tissier11,tarjus13,footnote_perturbRFIM}

An important issue concerning the above models is whether the critical behavior of the systems in and out of equilibrium, both being controlled by zero-temperature fixed points, are in the same universality class. For random elastic manifold models (which we will refer to as REMM in the following\cite{footnote_REMM}) it has been proven that this is not the case. The fixed point describing a random-bond pinning environment which is present in equilibrium disappears in the driven case at the depinning  threshold\cite{narayan92} and, more generally, it has been shown that the equilibrium and depinning fixed points are different when calculated at 2 loops of the perturbative functional renormalization group (FRG).\cite{FRGchauve,FRGledoussal-chauve} The question is therefore settled. In the case of the random-field Ising model (RFIM), there has been claims that the critical points in equilibrium and out-of-equilibrium (along the hysteresis loop) are the same,\cite{maritan94,perez04,liu09} but we have recently shown by using the nonperturbative FRG in a dynamical framework that this is not true:\cite{balog_eqnoneq} In- and out-of-equilibrium critical fixed points differ below the critical dimension $d_{DR}$ at which avalanches and cusps become relevant at large scale. However, because the analysis through the nonperturbative FRG is quite involved we find timely to check our entire procedure by applying it to the example of the REMM, for which we can repeat the very same steps as for the RFIM.

The purpose of the present work is to benchmark our nonperturbative FRG approach for in- and out-of-equilibrium criticality on the well-studied case of the REMM. We do not aim at uncovering some unknown physics about the model as its description is by now very well established. We want to check if the nonperturbative FRG as we implement it is able to capture the known results about the REMM and provide a quantitatively accurate description of its critical behavior in the pinned phase at equilibrium and at the depinning threshold in the athermally and quasi-statically driven protocol. As will be shown the answer is definitely positive, which gives strong support to our previous result concerning the distinct universality classes for the critical behavior of the RFIM in and out of equilibrium.

The outline of the paper is as follows. We first recall the dynamical field theory that describes an elastic manifold in a random environment (REMM) at large distance and long times for both the equilibrium situation and the athermally and quasi-statically driven protocol leading to a depinning transition. In the following section we review the nonperturbative FRG in the (dynamical) effective average action formalism and we derive the exact RG equations for the cumulants of the renormalized random force. We then introduce the nonperturbative approximation scheme and the minimal truncation required to capture the physics of the REMM in the equilibrium pinned phase and at the depinning transition. We next derive explicitly the NP-FRG flow equations for the second and the third cumulants of the random force as well as for the friction. We outline the distinct ways by which one handles the nonanalyticities appearing at the fixed points (the cusps in the functional dependence of the cumulants of the random force) for the equilibrium case, where the cusps can be regularized by an infinitesimal temperature, and in the driven case, where the cusps can be regularized by an infinitesimal velocity of the manifold. The procedure is similar to that used in the perturbative FRG treatment of the model\cite{FRGchauve,FRGledoussal-chauve} and is the exact analog of what we developed for treating the RFIM in and out of equilibrium.\cite{balog_eqnoneq} Finally we present the outcome, first the relation with the loop expansion and then the numerical results obtained for a whole range of dimensions, $0<d\leq 4$, and we compare these results with the known ones from the perturbative FRG and computer simulations. We briefly conclude and we complement the presentation by three appendices mostly discussing technical details.

\section{Dynamical field theory for an elastic manifold in a random environment}
\label{dynamical}

We consider an elastic manifold of internal dimension $d$ parametrized by a $1$-component displacement field (or height) $\varphi_x$, where $x$ denotes a $d$-dimensional space coordinate. The manifold is in a disordered environment and its equilibrium properties are described by a Hamiltonian that is the sum of an elastic term that tends to favor a flat manifold and a random potential $V(\varphi_x,x)$ that tends to pin the manifold in certain configurations:
\begin{equation}
\begin{aligned}
\label{eq_ham_dis_REMM}
&\mathcal H_{dis}[\varphi;V]=  \int_{x} \big\{\frac{1}{2}(\partial_x \varphi_x)^2+ V(\varphi_x,x)\big \} \,,
\end{aligned}
\end{equation}
where $ \int_{x} \equiv \int d^d x$. The random potential $V$ is taken with a Gaussian distribution characterized by a zero mean and  a variance $\overline{V(\varphi_1,x_1)V(\varphi_2,x_2)}= R_B(\varphi_1-\varphi_2) \delta^{(d)}( x_1-x_2)$ where the variance $R_B(\varphi)$ can take several functional forms depending on the physics of the problem at hand: $R_B(\varphi)$ is periodic for a system of pinned charge-density waves, is a short-range function when the disorder is of random-bond type, and is long-ranged, going as $-\vert \varphi \vert$ at large argument, for a random-field disorder. (Note that the latter case corresponds to the behavior of an interface in a random-field Ising model and should not be confused with the bulk behavior of the same model, which is what we studied in previous work.\cite{balog_eqnoneq}) An ultraviolet (UV) cutoff $\Lambda$ on the momenta, associated with the inverse of a microscopic length scale such as a lattice spacing, is also implicitly taken into account. All of this is by now very well known and has a long history of major contributions.\cite{fisher_dep,fisher86-pin,nattermann_dep,narayan92,FRGchauve,FRGledoussal-chauve,balents-ledoussal,ledoussal10,husemann18}

At a coarse-grained level, the dynamics of the system, near to and away from equilibrium, can be described by an overdamped Langevin equation,
\begin{equation}
\label{eq_stochastic_dynamics_REMM}
\eta_B\partial_t\varphi_{xt}=\partial_x^2 \varphi_{xt}+F(\varphi_{xt},x) + f_t+ \xi_{xt},
\end{equation}
where $\eta_B$ is the bare friction coefficient, $\xi_{xt}$ is a Gaussian random thermal noise with zero mean and variance $\langle\xi_{xt}\xi_{x't'}\rangle=2\eta_B T \delta^{(d)}(x-x')\delta(t-t')$,  $f_{t}$ is an applied force, and $F(\varphi,x)$ is a random pinning force with zero mean and variance $\overline{F(\varphi_1,x_1)F(\varphi_2,x_2)}= \Delta_B(\varphi_1-\varphi_2) \delta^{(d)}( x_1-x_2)$. At the bare level, one has $F(\varphi,x)=-\partial_\varphi V(\varphi,x)$ and $\Delta_B(\varphi)=-\partial_\varphi^2 R_B(\varphi)$. We are interested in the present work by two different situations: the relaxation to equilibrium in the pinned phase, which corresponds to taking $T>0$ and $f=0$, and  the quasi-statically driven situation leading to a depinning transition, which corresponds to $T=0$ and $f_t=f_c+\Omega t$ with $f_c$ the critical depinning threshold and $\Omega\to 0^+$. 

In the pinned phase at equilibrium and at the depinning threshold, the manifold is rough at large scale with
\begin{equation}
\label{eq_roughness}
\overline{(\varphi_{x_1}-\varphi_{x_2})^2} \sim \vert x_1-x_2 \vert ^{2\zeta} ,
\end{equation}
where $\zeta$ is the roughness exponent which has been shown to take different values in equilibrium and at the depinning threshold.\cite{FRGchauve,FRGledoussal-chauve}

The generating functional of the multi-point and multi-time correlation and response functions can be built as usual by following the Martin-Siggia-Rose-Janssen-de Dominicis formalism.\cite{MSR,janssen-dedom} We closely follow the steps of our work on the RFIM in and out of equilibrium but the derivation is very similar to previous studies of elatic manifolds in a random environment.\cite{nattermann_dep,narayan92,FRGledoussal-chauve,chauve-creep,balents-doussal_dyn} After introducing an auxiliary ``response'' field $\hat \varphi_{xt}$ and taking into account the fact that the solution of Eq. (\ref{eq_stochastic_dynamics_REMM}) is unique,\cite{zinnjustin89} one obtains the ``partition function''
\begin{equation}
\begin{aligned}
\label{eq_part_dis2}
\mathcal Z_{F,\xi}[\hat{f},f]=&\int \mathcal{D}\varphi\mathcal{D}\hat{\varphi} \exp\big\{-\int_{xt}\hat{\varphi}_{xt}\Big[ \eta_B \partial_t\varphi_{xt}-  \partial_x^2\varphi_{xt}  \\& 
-\xi_{xt} -F(\varphi_{xt},x)\Big]+\int_{xt}(\hat{f}_{xt}\varphi_{xt}+f_{xt}\hat{\varphi}_{xt})\Big \}\,
\end{aligned}
\end{equation}
where we have used the It\={o} prescription (which amounts to setting to 1 the Jacobian of the transformation between the thermal noise and the field).\cite{zinnjustin89} 

For handling the averages over the thermal noise and the random force we follow the same procedure as in Refs. [\onlinecite{balog_activated,balog_eqnoneq}]. To account for phenomena such as avalanches and droplets that are crucial to properly describe the physics of the present class of disordered models one needs to be able to keep track of the full functional dependence of the cumulants of the renormalized disorder,\cite{tarjus04,tissier06,tissier11} and to this aim we introduce copies or replicas of the system.  The copies have the {\it same} quenched disorder $F$ but are coupled to \textit{distinct} sources and are characterized by {\it independent} thermal noises (with the same Gaussian distribution). Contrary to the conventional practice,\cite{dedom78} we therefore combine  dynamics \textit{and} replicas.

After averaging over the thermal noises and the random force, we obtain
\begin{equation}
\begin{aligned}
\label{eq_part_aver}
Z[{\hat{f}_a,f_a}]= \int \prod_a \mathcal{D}\varphi_a\mathcal{D}\hat{\varphi}_a e^{-S_{dyn}[\{\hat{\varphi}_a,\varphi_a\}]+\sum_a\int_{xt}(\hat{f}_{a,xt}\varphi_{a,xt}+f_{a,xt}\hat{\varphi}_{a,xt})}
\end{aligned}
\end{equation}
with the bare dynamical action is given by
\begin{equation}
\begin{aligned}
\label{eq_bare_action}
&S_{dyn}[\{\hat{\varphi}_a,\varphi_a\}]= \\&
\sum_a \int_{xt}\hat{\varphi}_{a,xt}\Big\{\eta_B\left (\partial_t\varphi_{a,xt}-T\hat{\varphi}_{a,xt}\right ) -\partial_x^2\varphi_{a,xt}\Big\} -\frac{1}{2}\sum_{ab}\int_{xtt'}  \hat{\varphi}_{a,xt}\Delta_B(\varphi_{a,xt}-\varphi_{b,xt'})\hat{\varphi}_{b,xt'}\,.
\end{aligned}
\end{equation}
Note the difference between the effect of the thermal noises that lead to a one-replica term which is local both in space and in time and the effect of the quenched random force that generates a two-replica term which is local in space but not in time.

Causality comes with It\={o}'s prescription and should apply for both the equilibrium and the quasi-statically driven dynamics. In addition, the relaxation toward equilibrium satisfies an invariance under time translation and a time-reversal symmetry.\cite{zinnjustin89} The latter in turn implies the fluctuation-dissipation theorem relating correlation and response functions.\cite{janssen92,zinnjustin89,andreanov06} 
It applies only for $T>0$ and is therefore {\it a priori} violated in the out-of-equilibrium, driven case. The other symmetries of the dynamical action, besides translational and rotational invariance,  are a global $Z_2$ inversion symmetry, $\varphi_{a,xt} \to -\varphi_{a,xt}$, $\hat\varphi_{a,xt}\to -\hat\varphi_{a,xt}$, $\forall a$, and a ``statistical tilt symmetry'' (STS)\cite{shultz88,hwa-fisher94} under the displacement $\varphi_{a,xt} \to \varphi_{a,xt}+\chi_x$ in each replica with $\chi_x$ a replica- and time-independent field. The consequences of the STS and $Z_2$ symmetry will be further discussed below. In addition there is also an underlying supersymmetry, but it is effectively broken as soon as $d$ is less than the upper critical dimension $d_{uc}=4$\cite{wiese_SUSY} and, contrary to the case of  the RFIM,\cite{tissier11} we will not dwell on it.

\section{Nonperturbative functional renormalization group}
\label{NPFRG}

\subsection{The effective average action formalism}

As in our recent treatment of the RFIM in and out of equilibrium,\cite{balog_eqnoneq} we apply the nonperturbative functional renormalization group (NP-FRG) to describe the long-distance and long-time physics of the dynamical theory defined above. The exact renormalization group procedure relies on a progressive account of the contribution of the fluctuations of the fields on longer length and time scales, or alternatively, shorter momenta and frequencies.\cite{wilson74,wetterich93,berges02} The procedure can be implemented through the addition to the bare action of an ``infrared (IR) regulator'' $\Delta S_k$ depending on a running IR scale $k$. Its role is to suppress the integration over slow modes associated with momenta $\vert q \vert \lesssim k$ in the functional integral.\cite{wetterich93,berges02,tarjus04,tissier11} we therefore add to the replicated dynamical action in Eq. (\ref{eq_bare_action}) a quadratic term,
\begin{equation}
\begin{aligned}
\label{bare_action}
\Delta S_{k}[\{\Phi_a\}]= &\frac 12 \int_{xt}\int_{x't'}{\rm{tr}}\Big[\sum_a \Phi_{a,xt}\widehat{\mathbf R}_k(x-x',t-t') \Phi_{a,x't'}^\top \\&
+\frac 12 \sum_{ab} \Phi_{a,xt}\widetilde{\mathbf R}_k(x-x',t-t') \Phi_{b,x't'}^\top\Big] \,,
\end{aligned}
\end{equation}
where $\Phi_a\equiv (\varphi_a,\hat\varphi_a)$,  $\Phi_a^\top$ is its transpose, and the trace is over the 2 components of $\Phi_a$; $\widehat{\mathbf R}_k$ and $\widetilde{\mathbf R}_k$ are symmetric $2\times 2$ matrices of mass-like IR cutoff functions that enforce the decoupling between fast (high-momentum) and slow (low-momentum) modes in the partition function. Following previous work,\cite{canet,balog_eqnoneq} it proves sufficient to control the contribution of the fluctuations through their momentum dependence, and we take
\begin{equation}
\begin{aligned}
\label{hatR}
\widehat R_{k,11}=\widehat R_{k,22}=0 \;,  \widehat R_{k,12}=\widehat R_{k,21}=\widehat R_k(\vert x-x'\vert)\,,
\end{aligned}
\end{equation}
and 
\begin{equation}
\begin{aligned}
\label{tildeR}
\widetilde R_{k,11}=\widetilde R_{k,12}=\widetilde R_{k,21}=0 \;, \widetilde R_{k,22}=\widetilde R_k(\vert x-x'\vert)\,,
\end{aligned}
\end{equation}
where  the Fourier transforms, $\widehat R_k(q^2)$ and $\widetilde R_k(q^2)$, are chosen  such that the integration over modes with momentum $\vert q \vert \lesssim k$ is suppressed.\cite{berges02,tarjus04,tissier11} (Note that this choice of IR regulator easily satisfies the symmetries of the present theory.)

After adding the above IR regulator to the bare action, the partition function $Z[\{\mathcal F_a\}]$, where $\mathcal F_a$ denotes $(\hat f_a,f_a)$ is replaced by a $k$-dependent quantity,
\begin{equation}
\begin{aligned}
\label{eq_part_aver_k}
Z_k[\{\mathcal F_a\}]= \int \prod_a \mathcal{D}\Phi_a e^{-S_{dyn}[\{\Phi_a\}]+\sum_a\int_{xt} \mathcal{F}_{a,xt} \Phi_{a,xt}^\top-\Delta S_{k}[\{\Phi_a\}]}\,.
\end{aligned}
\end{equation}

The  central quantity of our NP-FRG is the ``effective average action" $\Gamma_k$,\cite{wetterich93} which is the generating functional of the 1-particle irreducible (1PI) correlation functions at the scale $k$. It is defined (modulo the subtraction of a regulator contribution) from $\ln Z_k[\{\mathcal F_a\}]$ via a Legendre transform:
\begin{equation}
\label{eq_legendre_transform_k}
\Gamma_k[\{\Phi_a\}]+\ln{Z_k[\{\mathcal F_a\}]}= \sum_a \int_{xt} {\rm{tr}} \mathcal{F}_{a,xt}\Phi_{a,xt}^\top-\Delta S_k[\{\Phi_a\}],
\end{equation}
where $\Phi_a\equiv (\phi_a,\hat\phi_a)$ now denotes the ``classical'' (or average) fields with 
\begin{equation}
\begin{aligned}
\label{eq_legendre}
&\phi_{a,xt}=\frac{\delta \ln Z_k}{\delta \hat f_{a,xt}}=\langle \varphi_{a,xt}\rangle
\\&
\hat{\phi}_{a,xt}=\frac{\delta \ln Z_k}{\delta f_{a,xt}}=\langle \hat\varphi_{a,xt}\rangle\,.
\end{aligned}
\end{equation}
The trace in Eq. (\ref{eq_legendre_transform_k}) is over the 2 components of $\Phi_a$ and $\mathcal F_a$. 

The effective average action $\Gamma_k$ satisfies an {\it exact} RG equation (ERGE) describing its evolution with the IR cutoff $k$,\cite{wetterich93}
\begin{equation}
\label{eq_wetterich}
\partial_k\Gamma_k[\{\Phi_a\}]=\frac 12 \rm{Tr} \big\{ (\partial_k\mathbf R_k) (\bm\Gamma_k^{(2)}[\{\Phi_a\}] + \mathbf R_k)^{-1} \big\},
\end{equation}
where the trace is over space-time coordinates, copy indices and components, and $\bm\Gamma_k^{(2)}$ is the matrix formed by the second functional derivatives of $\Gamma_k$. (In what follows, superscripts within a parenthesis are used to indicate derivatives with respect to the appropriate arguments.)

\subsection{ERGE's for the cumulants of the renormalized random force}

We are interested in the cumulants of the renormalized disorder, here of the renormalized random force. The cumulants at the scale $k$ can be generated by expanding the effective average action in increasing number of unrestricted sums over copies,\cite{tarjus04,tissier06,tissier11,balog_eqnoneq,ledoussal10}
\begin{equation}
\label{eq_expansion_gamma_k}
\Gamma_k[\{\Phi_a\}]=\sum_{p=1}^\infty \frac {(-1)^{p-1}}{p!}\sum_{a_1\cdots a_p} \mathsf \Gamma_{kp}[\Phi_{a_1},\cdots,\Phi_{a_p}]\,,
\end{equation}
where the (generalized) cumulants $\mathsf \Gamma_{kp}$ are invariant under any permutation of their $p$ arguments. As a result of causality and It\=o's prescription,\cite{balog_eqnoneq} $\mathsf \Gamma_{kp}$ can be cast in the following form:
\begin{equation}
\begin{aligned}
\label{eq_gamma_kp}
\mathsf \Gamma_{kp}=\int_{x_1t_1}\cdots \int_{x_p t_p}\hat\phi_{a_1,x_1t_1}\cdots \hat\phi_{a_p,x_pt_p} \gamma_{k p;x_1t_1,\cdots,x_pt_p}
\end{aligned}
\end{equation}
where $\gamma_{k p;x_1t_1,\cdots,x_pt_p}$ is a functional of the fields  $\Phi_{a_1,t_1},\cdots, \Phi_{a_p,t_p}$ and of their time derivatives, $\partial_{t_1}^q \Phi_{a_1,t_1},\cdots, \partial_{t_p}^q \Phi_{a_p,t_p} $, $q\geq 1$. Note that the fields and their time derivatives are taken at fixed time values $t_1,\cdots,t_p$ whereas space points are not specified. The properties of the $\gamma_{k p}$'s are discussed in more detail in Ref.~[\onlinecite{balog_eqnoneq}] (see also [\onlinecite{balents-doussal_dyn}]).



By differentiating twice the cumulant expansion in Eq.~(\ref{eq_expansion_gamma_k}), one has access to the expansion of the 2-point 1-PI correlation functions (or proper vertices). The components of the propagator $\mathbf P_{k}=(\bm{\Gamma}_k^{(2)}+ \mathbf R_k)^{-1}$, {\it i.e.},  $\mathbf P_{k,ab}=\delta_{ab} \widehat{\mathbf P}_{k,a} + \widetilde{\mathbf P}_{k,ab}$, can also be expanded in increasing number of unrestricted sums over copies and, through a term-by-term identification, related to second derivatives of the cumulants. From the lowest order, one obtains
\begin{equation}
\label{eq_hatP_zero}
\mathbf{\widehat {P}}_{k;x_1t_1,x_2t_2}^{[0]}[\Phi ]=\left(\bm{\mathsf\Gamma}_{k1}^{(2)}[ \Phi ]+\mathbf{\widehat R}_k\right) ^{-1}\Big \vert_{x_1t_1,x_2t_2}
\end{equation}
and
\begin{equation}
\begin{aligned}
\label{eq_tildeP_zero}
&\mathbf{\widetilde {P}}_{k;x_1t_1,x_2t_2}^{[0]}[\Phi_1, \Phi_2 ]= \int_{x_3t_3}\int_{x_4t_4}\mathbf{\widehat {P}}_{k;x_1t_1,x_3t_3}^{[0]}[ \Phi_1 ] \times \\&
\Big (\bm{\mathsf\Gamma}_{k2;x_3t_3,x_4t_4}^{(11)}[\Phi_1, \Phi_2 ]
- \mathbf{\widetilde R}_k(\vert x_3-x_4\vert)\Big ) \mathbf{\widehat {P}}_{k;x_4t_4,x_2t_2}^{[0]}[ \Phi_2 ]\,.
\end{aligned}
\end{equation}

Finally, after inserting Eq.~(\ref{eq_expansion_gamma_k}) into the ERGE for the effective average action, Eq.~(\ref{eq_wetterich}),  and after some algebraic manipulations, one can derive an infinite hierarchy of coupled ERGE's for the generalized cumulants $\mathsf \Gamma_{kp}$. For instance, with the choice of IR regulator in Eqs.~(\ref{hatR},\ref{tildeR}) and after setting for simplicity the IR cutoff function $\widetilde {R}_k$ to zero (see the discussion further below), the first three flow equations explicitly read
\begin{equation}
\label{eq_flow_Gamma1}
\begin{aligned}
\partial_k \mathsf \Gamma_{k 1}\left[ \Phi_1\right ]=
\dfrac{1}{2} \int_{x_1x_2}  \int_{t_1}{\rm{tr}} \Big [\partial_k \widehat{\mathbf R}_k(\vert x_1-x_2\vert) \big (\widehat{\mathbf P}_{k;(x_1 t_1)(x_2 t_1)}^{[0]}[ \Phi_1] +\widetilde{\mathbf P}_{k;(x_1t_1)(x_2 t_1)}^{[0]}[ \Phi_1, \Phi_1] \big )
\Big ]
\end{aligned}
\end{equation}
\begin{equation}
\label{eq_flow_Gamma2}
\begin{aligned}
&\partial_k \mathsf \Gamma_{k2}\left[ \Phi_1 , \Phi_2\right ]= \dfrac{1}{2} \widetilde{\partial}_k \bigg \{ \int_{x_3x_4} \int_{t_3t_4} {\rm{tr}}
 \Big [ \widehat{\mathbf P}_{k;(x_3 t_3)(x_4 t_4)}^{[0]}\left[ \Phi_1 \right ] \times
 \\& ( \bm{\mathsf \Gamma} _{k2;(x_4 t_4)(x_3 t_3),.}^{(20)}\left[ \Phi_1,  \Phi_2 \right ] - \bm{\mathsf \Gamma} _{k3;(x_4 t_4),(x_3 t_3),.}^{(110)}\left[ \Phi_1,  \Phi_1,  \Phi_2 \right ])  
\\& + \widetilde{\mathbf P}_{k;(x_3t_3) (x_4t_4)}^{[0]}\left[ \Phi_1, \Phi_1 \right ] 
\bm{\mathsf \Gamma}_{k2;(x_4 t_4)(x_3 t_3),.}^{(20)}\left[ \Phi_1, \Phi_2 \right ] 
 + \\&
 \dfrac{1}{2} \widetilde{ \mathbf P}_{k;(x_3t_3)(x_4 t_4)}^{[0]}\left [ \Phi_1, \Phi_2 \right ]\bm{\mathsf \Gamma}_{k2;(x_4 t_4),(x_3 t_3)}^{(11)}\left[ \Phi_1, \Phi_2 \right ]  + perm (12) \Big ]\bigg \} \,,
\end{aligned}
\end{equation}
\begin{equation}
\label{eq_flow_Gamma3}
\begin{aligned}
&\partial_k \mathsf \Gamma_{k3}\left[ \Phi_1 , \Phi_2, \Phi_3\right ]= \dfrac{1}{2} \widetilde{\partial}_k \bigg \{ \int_{x_1 x_2} \int_{t_1 t_2} {\rm{tr}}
 \Big [\frac{1}{2}\widehat{\mathbf P}_{k;(x_1 t_1)(x_2 t_2)}^{[0]}\left[ \Phi_1 \right ] \times
 \\&
 (\bm{\mathsf \Gamma} _{k3;(x_2 t_2)(x_1 t_1),.,.}^{(200)}\left[ \Phi_1,  \Phi_2,  \Phi_3 \right ] - \bm{\mathsf \Gamma} _{k4;(x_2 t_2),(x_1 t_1),.,.}^{(1100)}\left[ \Phi_1,  \Phi_1,  \Phi_2,  \Phi_3 \right ])
 \\&
 +\frac{1}{2}\widetilde{\mathbf P}_{k;(x_1t_1) (x_2t_2)}^{[0]}\left[ \Phi_1, \Phi_1 \right ]\bm{\mathsf \Gamma} _{k3;(x_2 t_2)(x_1 t_1),.,.}^{(200)}\left[ \Phi_1,  \Phi_2,  \Phi_3 \right ] +
 \\&
 \widetilde{\mathbf P}_{k;(x_1t_1) (x_2t_2)}^{[0]}\left[ \Phi_1, \Phi_2 \right ]\bm{\mathsf \Gamma} _{k3;(x_2 t_2)(x_1 t_1),.}^{(110)}\left[ \Phi_2,  \Phi_1,  \Phi_3 \right ] + perm (123) \Big ]
 \\& 
 + \int_{x_1 x_2 x_3 x_4} \int_{t_1 t_2 t_3 t_4} {\rm{tr}} \Big [\widehat{\mathbf P}_{k;(x_1 t_1)(x_2 t_2)}^{[0]}\left[ \Phi_1 \right]\bm{\mathsf \Gamma} _{k2;(x_2 t_2)(x_3 t_3),.}^{(20)}\left[ \Phi_1,  \Phi_2 \right]\widehat{\mathbf P}_{k;(x_3 t_3)(x_4 t_4)}^{[0]}\left[ \Phi_1 \right]
 \\&
 \times (\bm{\mathsf \Gamma} _{k3;(x_4 t_4),(x_1 t_1),.}^{(110)}\left[ \Phi_1,  \Phi_1,  \Phi_3 \right ] -\frac{1}{2} \bm{\mathsf \Gamma} _{k2;(x_4 t_4)(x_1 t_1),.}^{(20)}\left[\Phi_1,  \Phi_3 \right ])
 \\&
 -\widehat{\mathbf P}_{k;(x_1 t_1)(x_2 t_2)}^{[0]}\left[ \Phi_1 \right]\bm{\mathsf \Gamma} _{k2;(x_2 t_2)(x_3 t_3),.}^{(20)}\left[ \Phi_1,  \Phi_2 \right]\times 
 \\&
 (\widetilde{\mathbf P}_{k;(x_3t_3) (x_4t_4)}^{[0]}\left[ \Phi_1, \Phi_1 \right ]\bm{\mathsf \Gamma} _{k2;(x_4 t_4)(x_1 t_1),.}^{(20)}\left[\Phi_1,  \Phi_3 \right ]+\widetilde{\mathbf P}_{k;(x_3t_3) (x_4t_4)}^{[0]}\left[ \Phi_1, \Phi_3 \right ]\bm{\mathsf \Gamma} _{k2;(x_4 t_4),(x_1 t_1)}^{(11)}\left[\Phi_3,  \Phi_1 \right ])
 \\&
 -\widetilde{\mathbf P}_{k;(x_1 t_1)(x_2 t_2)}^{[0]}\left[ \Phi_1,\Phi_2 \right]\bm{\mathsf \Gamma}_{k2;(x_2 t_2),(x_3 t_3)}^{(11)}\left[ \Phi_2,  \Phi_3 \right]\widehat{\mathbf P}_{k;(x_3t_3) (x_4t_4)}^{[0]}\left[ \Phi_3\right]\bm{\mathsf \Gamma}_{k2;(x_4 t_4),(x_1 t_1)}^{(11)}\left[\Phi_3,  \Phi_1 \right ]  
 \\&
 + perm (123) \Big] \bigg \} ,
\end{aligned}
\end{equation}
where we recall that the superscripts within parentheses on the $\mathsf \Gamma_{k p}$'s indicate functional derivatives. In the second and third equations we have introduced the short-hand notation $\widetilde{\partial}_k$ to indicate a derivative acting only on the cutoff functions (\textit{i.e.},  $\widetilde{\partial}_k \equiv \partial_k \widehat{R}_k\, \delta/\delta \widehat{R}_k$) and $perm (12)$ and $perm(123)$ denote the expressions obtained by permuting $ \Phi_1$ and $ \Phi_2$ or $ \Phi_1$, $ \Phi_2$ and $ \Phi_3$ (respectively). Finally, the trace $ {\rm{tr}} [\,]$ is over the components of the $2\times 2$ matrices.

\section{Nonperturbative approximation scheme}
\label{approx}

The hierarchy of ERGE's cannot be solved exactly and we consider the same nonperturbative approximation scheme for the effective average action that we used for studying the RFIM in and out of equilibrium.\cite{balog_eqnoneq} It combines a truncation in the \textit{spatial derivative expansion}, \textit{i.e.}, an expansion in the number of spatial derivatives for approximating the long-distance behavior of the 1PI correlation functions, a truncation in the  \textit{time derivative expansion and expansion in the auxiliary fields $\hat \phi_a$}, which amounts to truncating the number of kinetic coefficients for describing the long-time behavior,\cite{balents-doussal_dyn,gorokhov_creep} and a truncation in the \textit{expansion in cumulants of the renormalized disorder}. 

In the case of the RFIM, a truncation after the second cumulant is already sufficient to capture all the physics of the problem and distinguish between equilibrium and out-of-equilibrium critical behavior.\cite{balog_eqnoneq} {\it A priori} indeed, there is a symmetry difference between the equilibrium situation where the critical point takes place in zero sources (zero magnetic field) and the driven case where the critical points along the hysteresis curves take place for nonzero values of the magnetic field: In the latter case the $Z_2$ symmetry of the bare dynamical action may be broken at the fixed point whereas it is satisfied in the former case, and this can be seen in the functional dependence of the first two cumulants. 

In the random elastic manifold model (REMM) there is also a potential symmetry difference between equilibrium, which takes place for a zero applied force, and depinning, which takes place for a nonzero value of the force. However this breaking or not of the $Z_2$ symmetry at the fixed point can no longer be easily detected by working with only the first two cumulants and neglecting all higher-order ones.  This is due to the strong constraints (and simplifications!) resulting from the STS. The latter indeed implies that the first cumulant $\gamma_{k1;x_1t_1}$ (defined in Eq.~(\ref{eq_gamma_kp}) with $p=1$), when evaluated for $\phi_{a_1,x_1t_1}$ independent of time and $\hat\phi_{a_1,x_1t_1}=0$,  is not renormalized, {\it i.e.}, $\gamma_{k1;x_1t_1}=-\partial_{x_1}^2\phi_{a_1,x_1}$ at all scale $k$, as in the bare dynamical action. In addition, the higher-order cumulants are invariant through a translation of all the replica fields $\phi_{a,xt}$ by a replica- and time-independent field $\chi_x$ at the same point $x$, {\it i.e.}, for $p\geq 2$,
\begin{equation}
\label{eq_STS}
\mathsf \Gamma_{kp}[(\phi_{a_1},\hat \phi_{a_1})\cdots,(\phi_{a_p},\hat \phi_{a_p})]=\mathsf \Gamma_{kp}[(\phi_{a_1}+\chi,\hat \phi_{a_1}),\cdots,(\phi_{a_p}+\chi,\hat \phi_{a_p})]\,.
\end{equation}
For instance, for replica fields that are independent of time with $\hat\phi_a=0$, the second cumulant $\gamma_{k2}$ (defined in Eq.~(\ref{eq_gamma_kp}) with $p=2$) is an even function of the field difference $\phi_{1,x}-\phi_{2,x}$ due to STS and permutation invariance, whether or not the $Z_2$ symmetry is broken.

The minimal truncation that already contains the long-distance and long-time physics of the REMM both in and out of equilibrium, and is able to predict a difference of behavior between the two situations when there is one, can then be formulated as
\begin{equation}
\begin{aligned}
\label{eq_gammak1_ansatz}
\mathsf \Gamma_{k 1}[\Phi]=&\int_{xt}\hat\phi_{xt}\Big [-\partial_x^2\phi_{xt}  + \eta_k\left (\partial_t\phi_{xt}-T \hat{\phi}_{xt}\right )\Big ]
\end{aligned}
\end{equation}
\begin{equation}
\begin{aligned}
\label{eq_gammak2_ansatz}
\mathsf{\Gamma}_{k2}[\Phi_1,\Phi_2]=\int_{x}\int_{t_1t_2}\hat{\phi}_{1,xt_1}\hat{\phi}_{2,xt_2}\Delta_k(\phi_{1,xt_1},\phi_{2,xt_2})
\end{aligned}
\end{equation} 
\begin{equation}
\begin{aligned}
\label{eq_gammak3_ansatz}
\mathsf{\Gamma}_{k3}[\Phi_1,\Phi_2,\Phi_3]=\int_{x}\int_{t_1t_2t_3}\hat{\phi}_{1,xt_1}\hat{\phi}_{2,xt_2}\hat{\phi}_{3,xt_3}W_k(\phi_{1,xt_1},\phi_{2,xt_2},\phi_{3,xt_3})
\end{aligned}
\end{equation} 
\begin{equation}
\begin{aligned}
\label{eq_gammak4_ansatz}
\mathsf{\Gamma}_{kp}=0,\; \forall p\geq 4\,.
\end{aligned}
\end{equation} 
The only approximation at the level of the first cumulant is that we have kept a renormalized friction term $\eta_k$ but neglected higher-order kinetic coefficients associated with higher time derivatives of $\phi_{xt}$ and higher powers of the response field $\hat\phi_{xt}$. The nonrenormalization of the coefficient in front of $\partial_x^2\phi_{xt}$ and the absence of powers of $\phi_{xt}$ or of higher spatial gradient terms are a consequence of STS. In equilibrium, the same friction coefficient appears in front of the two terms in $\partial_t\phi_{xt}$ and $T \hat{\phi}_{xt}$ as a result of the time-reversal symmetry and fluctuation-dissipation theorem. In the driven athermal case, the term proportional to temperature is equal to zero so that, again, only one friction coefficient is necessary. At the level of the second and third cumulant we have used a local approximation that focuses on the uniform (zero-momentum and zero-frequency) behavior. 

\begin{figure}[h!]
  \begin{center}
   \includegraphics[width=0.5\linewidth]{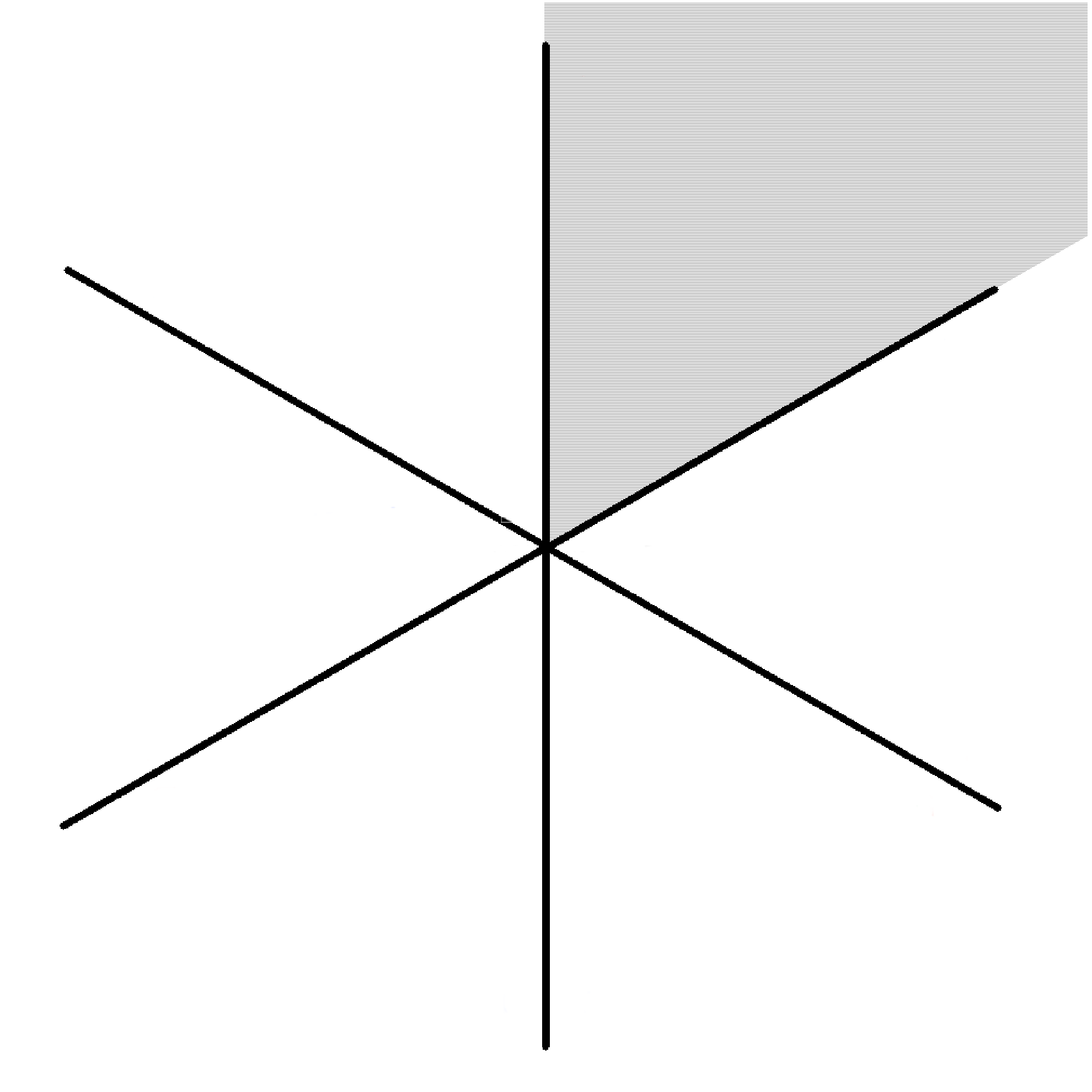}\includegraphics[width=0.5\linewidth]{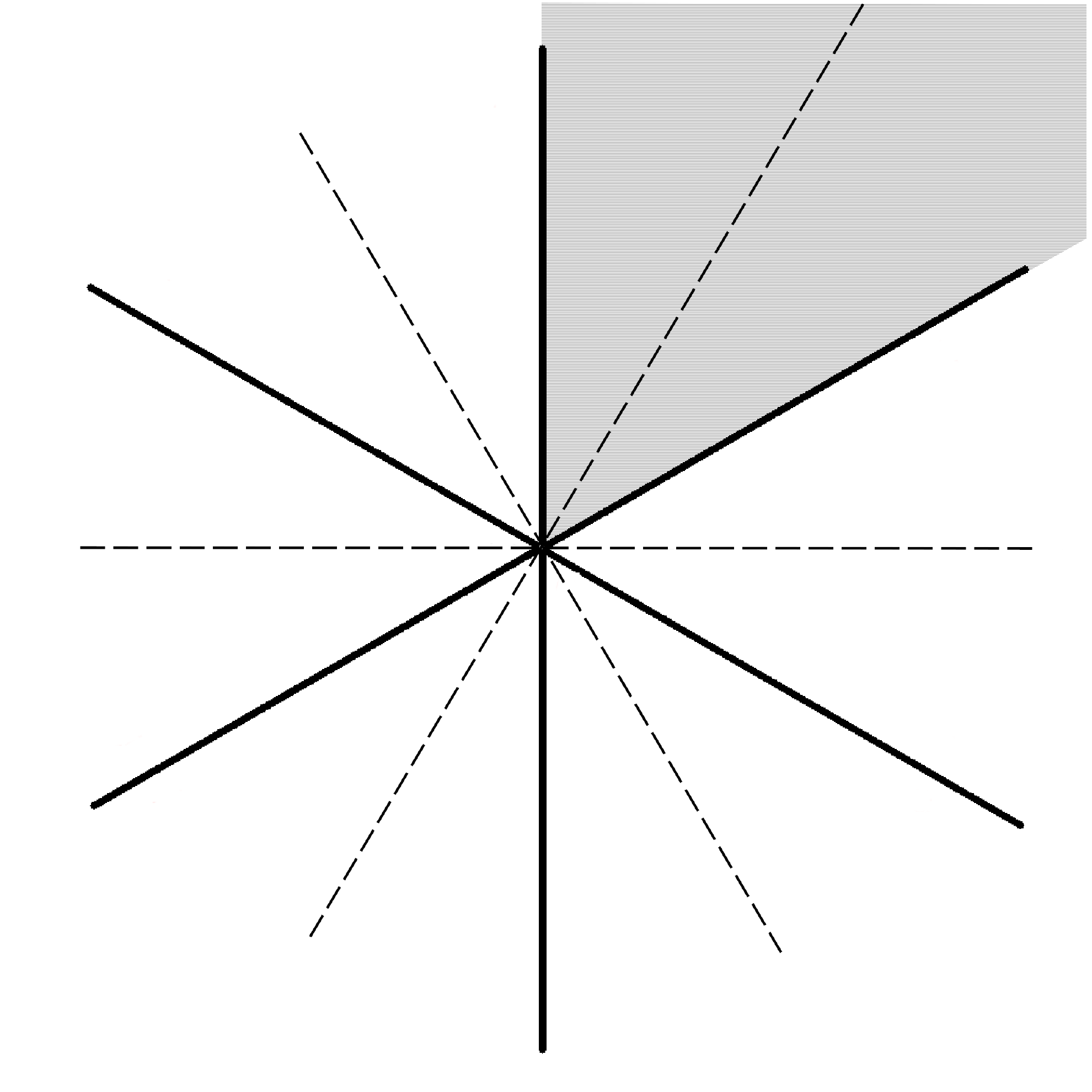}
    \caption{Representation of the behavior of the third cumulant $W_k(\phi_{1},\phi_{2},\phi_{3})\equiv W_k(y=(\phi_2-\phi_1)/\sqrt 2,  
z=(\phi_1+\phi_2-2\phi_3)/\sqrt 6)$ in the $(y,z)$ plane. The left panel correspond to the depinning case ($f=f_c^+$) and the right panel to the equilibrium case (zero applied force, $f=0$). In both cases there is an invariance of the function $W_k$ under $2\pi/3$ rotations (about the origin) and mirror symmetries about the three axes shown as full lines in the two panels. In the equilibrium case there is an additional invariance under a $\pi/3$ rotation followed by a change of sign: The associated axes of anti-symmetry are shown as dashed lines in the right panel. (In this case one also has that $W_k(0,0)=0$.) Because of the symmetries, the functional dependence of the third cumulant can be restricted to the shaded triangular section (or any equivalent one obtained by a $2\pi/3$ rotation).}\label{fig_w_dep}
  \end{center}
\end{figure}

The above truncation is very much in the spirit of that we have used to treat the RFIM,\cite{balog_eqnoneq} except that the drastic simplifications brought by the STS force us to move up one order in the truncation of the cumulant expansion. Note however that there is no associated increase in complexity as, due to STS and permutation invariance, the function $\Delta_k$ only depends on one argument and the function $W_k$ on two. To see this more explicitly, it is convenient to parametrize the field dependence by introducing
\begin{equation}
\begin{aligned}
\label{eq_field_parametrization}
&y=\frac{\phi_2-\phi_1}{\sqrt 2} \\&
z=\frac{\phi_1+\phi_2-2\phi_3}{\sqrt 6}\,.
\end{aligned}
\end{equation}
Then, one finds that 
\begin{equation}
\begin{aligned}
\label{eq_symmetries}
&\Delta_k(\phi_1,\phi_2)\equiv \Delta_k(y)=\Delta_k(-y) \\&
W_k(\phi_{1},\phi_{2},\phi_{3})\equiv W_k(y,z)=W_k(-y,z)=W_k(-\frac y2 -\frac{\sqrt 3z}2 ,\frac{\sqrt 3y}2-\frac z2)\,.
\end{aligned}
\end{equation}
The above symmetries allow us to restrict study of $\Delta_k(y)$ for $y\geq 0$ and that of $W_k(y,z)$ to a triangular section of the $(y,z)$ plane, {\it e.g.}, $y\geq 0, z\geq \sqrt 3 y/3$. This is pictorially represented in Fig.~\ref{fig_w_dep}. 

In the case of equilibrium, there is an additional symmetry resulting for the original global $Z_2$ symmetry in zero applied force,
\begin{equation}
\begin{aligned}
\label{eq_symmetries_eq}
W_k(y,z)=-W_k(y,-z)=-W_k(-y,-z)\,,
\end{aligned}
\end{equation}
which in particular implies that $W_k(0,0)=0$.

Finally, the renormalized quantities appearing in Eqs.~(\ref{eq_gammak1_ansatz}-\ref{eq_gammak3_ansatz}) are defined from the following prescriptions:
\begin{equation}
\begin{aligned}
 \label{eq_prescriptions}
 &\eta_k=\lim_{\omega\to 0}\frac{d}{d(-i\omega)} FT_{\omega} \frac{\delta^2\mathsf{\Gamma}_{k1}[\Phi]}{\delta\phi_{xt}\delta\hat{\phi}_{xt'}}\Big\vert_{unif}\\&
 \Delta_k(\phi_{1},\phi_{2})\delta^{(d)}(x_1-x_2)=\frac{\delta^2\mathsf{\Gamma}_{k2}[\Phi_1,\Phi_2]}{\delta\hat{\phi}_{1,xt_1}\delta\hat{\phi}_{2,xt_2}}\Big\vert_{unif}
 \\&
  W_k(\phi_{1},\phi_{2},\phi_{3})\delta^{(d)}(x_1-x_2)\delta^{(d)}(x_1-x_3)=
  \frac{\delta^3\mathsf{\Gamma}_{k3}[\Phi_1,\Phi_2,\Phi_3]}{\delta\hat{\phi}_{1,xt_1}\delta\hat{\phi}_{2,xt_2}\delta\hat{\phi}_{3,xt_3}}\Big\vert_{unif}\,.
\end{aligned}
\end{equation}
where $FT_{\omega}$ denotes a Fourier transform and the subscript $unif$ means that we take configurations of the fields that are spatially uniform with $\phi_{a,xt}=\phi_{a,t}$ and $\hat\phi_{a,xt}=0$. We will consider configurations of the physical fields $\phi_a$ that are also either strictly uniform, {\it i.e.} constant, in time (for equilibrium) or very slowly evolving in an appropriate quasi-static limit (for depinning). Indeed, relaxation to equilibrium corresponds to taking $T>0$ and the driving rate of the applied force $\Omega=0$. Due to the time reversal symmetry and the resulting fluctuation-dissipation theorem, the static quantites are independent of the dynamical ones. The limit to $T=0$ can then be taken and requires a careful account of the nonuniform convergence associated with the presence of a thermal boundary layer.\cite{chauve_creep,balents-ledoussal,ledoussal10,tissier06,tissier11,balog_activated}  On the other hand, the depinning transition corresponds to first setting $T=0$ and then considering the quasi-static limit of $\Omega\to 0^+$. This infinitesimal driving rate for $f \to f_c^+$ translates into an infinitesimal velocity for the field, so that we have to  consider configurations such that\cite{FRGledoussal-chauve}
\begin{equation}
\label{eq_driving_velocity}
\phi_t=\phi + vt,\; v\to 0^+\,.
\end{equation}

After inserting Eqs.~(\ref{eq_gammak1_ansatz}-\ref{eq_gammak4_ansatz}) in the ERGE's, Eqs.~(\ref{eq_flow_Gamma1},\ref{eq_flow_Gamma2},\ref{eq_flow_Gamma3}), and using the above RG prescriptions, one obtains a closed set of coupled nonperturbative functional RG equations for the functions $\Delta_k$, $W_k$ and the friction coefficient $\eta_k$. We again follow closely the steps of our previous study of the RFIM.\cite{balog_eqnoneq}

\section{NP-FRG flow equations}
\label{sec_NPFRG}

\subsection{Hat and tilde propagators}
\label{propagators}

Before studying in more detail the flow equations for $\eta_k$,  $\Delta_k(y)$, and $W_k(y,z)$ we first consider the propagators appearing in these equations. In all of these equations, the ``hat propagator''  matrix $\mathbf{\widehat {P}}_{k}^{[0]}[\Phi ]$ defined in Eq.~(\ref{eq_hatP_zero}) appears in field configurations that are uniform in space with moreover $\hat\phi=0$. It can be obtained from the second derivative of the approximate first cumulant expression given in Eq. (\ref{eq_gammak1_ansatz}). It is found to be independent of the field and simply given by
\begin{equation}
\begin{aligned}
\label{hatpropagator}
&\mathbf{\widehat {P}}_{k}^{[0]}(q^2,\omega)=  
\Bigg( \begin{array}{cc}
G_k(q^2,\omega) \;& \; G_k^+(q^2,\omega)\\
\\
G_k^-(q^2,\omega) \; &\; 0 \end{array} \Bigg)
\end{aligned}
\end{equation}
where, after transforming back to time, the response function at scale $k$ $G_k^+$  reads
\begin{equation}
\begin{aligned}
\label{response_function}
G_k^+(q^2,t',t)=\frac 1{\eta_k} e^{-\frac{q^2+\widehat R_k(q^2)}{\eta_k}(t-t')}\theta(t-t')\,,
\end{aligned}
\end{equation}
with $\theta$ the Heaviside step function,  $G_k^-(q^2,t',t)=G_k^+(q^2,t,t')$, and the correlation function at scale $k$ $G_k$ is equal to
\begin{equation}
\begin{aligned}
G_k(q^2,t',t) =T \,\frac {e^{-\frac{q^2+\widehat R_k(q^2)}{\eta_k}\vert t-t' \vert}}{q^2+\widehat R_k(q^2)}
\end{aligned}
\end{equation}
and is equal to zero when $T=0$.

It\=o's prescription is enforced in the nonperturbative RG by ensuring that the response function $G_k^+(t',t)$ is zero when the two times coincide. This is achieved by everywhere shifting the time for the auxiliary response field by an infinitesimal positive amount:  $\langle\hat\varphi_{x't'}\varphi_{xt} \rangle_k \to \langle\hat\varphi_{x't'+\epsilon}\varphi_{xt} \rangle_k$ with $\epsilon \to 0^+$.\cite{canet} This guarantees causality.

We now turn to the ``tilde propagator'' matrix $\mathbf{\widetilde {P}}_{k}^{[0]}[\Phi ]$. From Eq.~(\ref{eq_tildeP_zero})  and the  ansatz in Eq.~(\ref{eq_gammak2_ansatz}), it is easy to see that when evaluated for uniform configurations with the response fields set to zero the ``tilde propagator'' matrix  has only one nonzero component, the upper left one, which is then simply given in Fourier space by 
\begin{equation}
\begin{aligned}
\label{tildepropagator}
[\mathbf{\widetilde {P}}_{k}^{[0]}]_{11}(q^2,t_1,t_2;\phi_1,\phi_2)=  G_k^+(q^2,\omega=0)^2[\Delta_k(\phi_{1},\phi_{2})-\widetilde R_k(q^2)]\,, 
\end{aligned}
\end{equation}
which is purely static.

In what follows we will set for simplicity $\widetilde R_k=0$. This does not prevent suppressing fluctuations on scales smaller than the IR cutoff $k$, as we keep the cutoff function $\widehat R_k$. As we showed in the case of the RFIM,\cite{tissier11} $\widetilde R_k$ is important to ensure that the underlying supersymmetry of the field-theoretical construction which leads to dimensional reduction is not explicitly broken. However in the present case, it has been shown that this supersymmetry is broken as soon as $d<d_{uc}=4$ and that the results strongly deviate from the dimensional reduction predictions.\cite{wiese_SUSY} Neglecting $\widetilde R_k$ is therefore expected to be rather benign.

\subsection{Flow equations for the second and third cumulants in a graphical representation}

From the procedure outlined above and in our previous treatment of the RFIM criticality in and out of equilibrium,\cite{balog_eqnoneq} we obtain the nonperturbative FRG equations\cite{footnote_nonperturbative} for the second and third cumulants of the renormalized random force which can be expressed for both equilibrium and depinning in the following graphical form:
\begin{eqnarray}
\label{eq_2nd}
&\partial_k\Delta_k(\phi_{1t_1},\phi_{2t_2})=-\frac{\tilde{\partial}_k}{2}\int_q\int_{t_3t_4}\Bigg(\raisebox{-15pt}{\includegraphics[height=30pt,keepaspectratio]{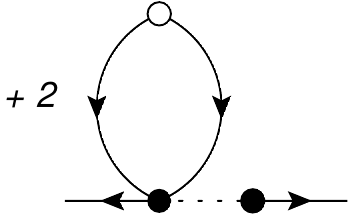}}\nonumber\\
&\raisebox{-15pt}{\includegraphics[height=30pt,keepaspectratio]{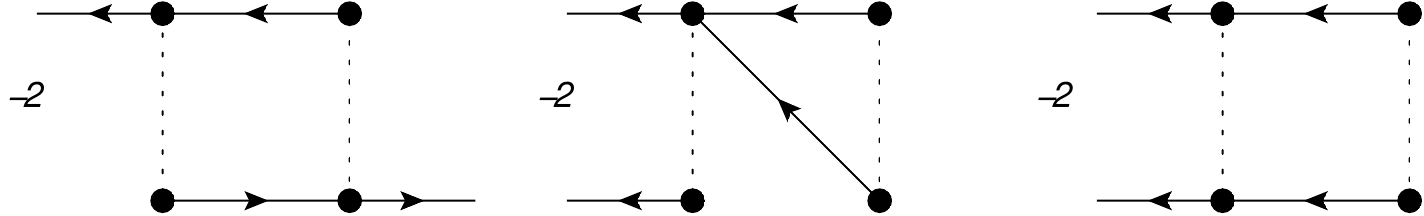}}\nonumber\\
&\raisebox{-15pt}{\includegraphics[height=35pt,keepaspectratio]{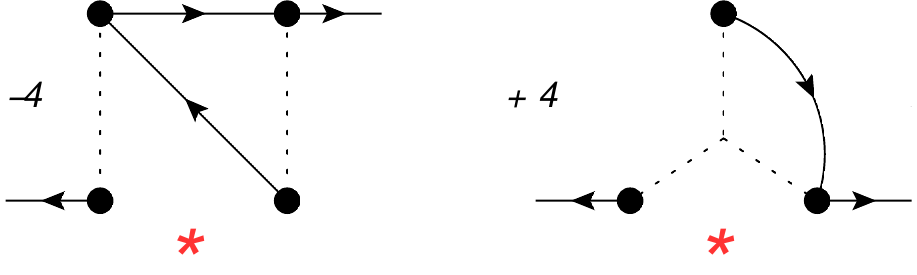}}\Bigg).
\end{eqnarray}

\begin{eqnarray}
\label{eq_3rd}
&\partial_k W_k(\phi_{1t_1},\phi_{2t_2},\phi_{3t_3})=\frac{\tilde{\partial}_k}{2}\int_q\int_{t_4t_5t_6}\Bigg(\raisebox{-15pt}{\includegraphics[height=30pt,keepaspectratio]{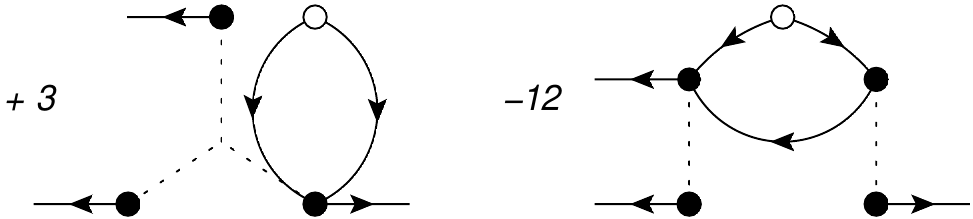}}\nonumber\\
&\raisebox{-15pt}{\includegraphics[height=30pt,keepaspectratio]{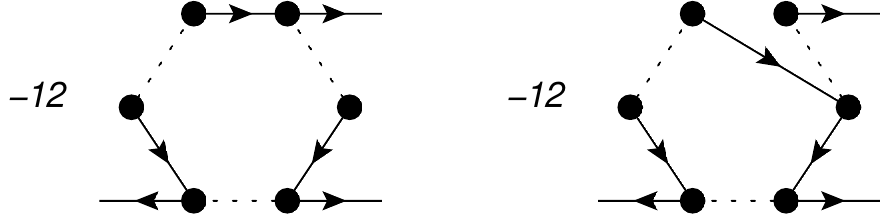}}\nonumber\\
&\raisebox{-15pt}{\includegraphics[height=30pt,keepaspectratio]{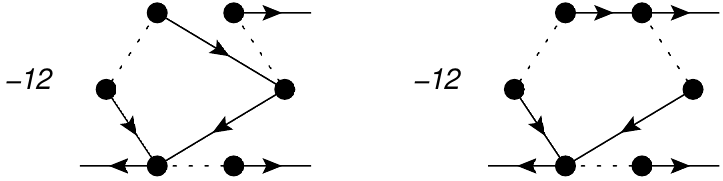}}\nonumber\\
&\raisebox{-15pt}{\includegraphics[height=30pt,keepaspectratio]{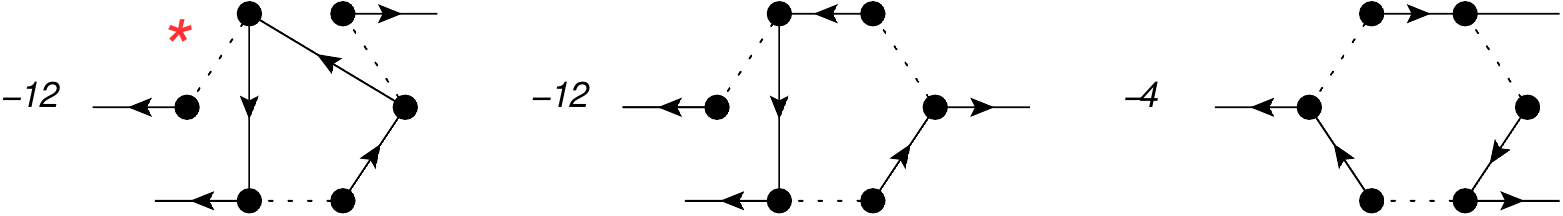}}\nonumber\\
&\raisebox{-15pt}{\includegraphics[height=30pt,keepaspectratio]{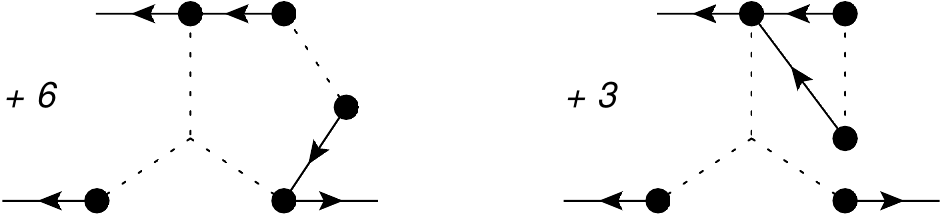}}\nonumber\\
&\raisebox{-15pt}{\includegraphics[height=30pt,keepaspectratio]{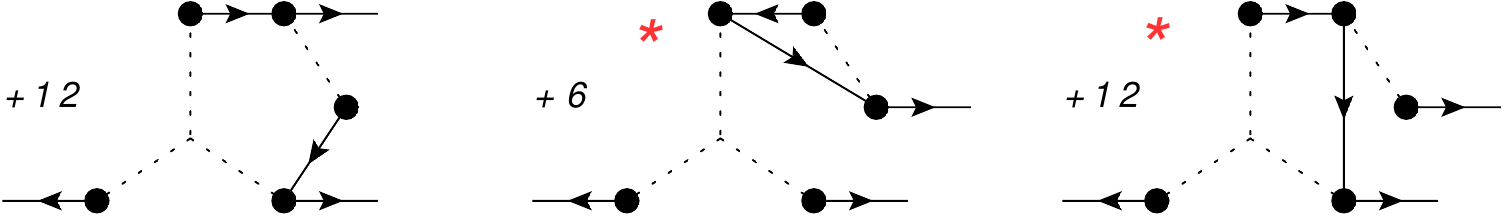}}\nonumber\\
&\raisebox{-15pt}{\includegraphics[height=30pt,keepaspectratio]{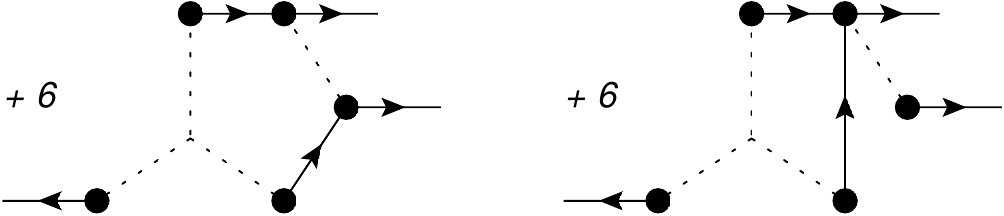}}\Bigg) \,, 
\end{eqnarray} 
where we have used the shorthand notations $\int_q \equiv \int d^dq/(2\pi)^d$ and $\int_{t}=\int_{-\infty}^{+\infty}dt$.

The propagator (response function) is represented graphically as
\begin{equation}
\label{gp}
G_k^+(q^2,t,t';\phi)=\raisebox{-10pt}{\includegraphics[width=60pt,keepaspectratio]{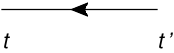}}
\end{equation}
and the correlation function as
\begin{equation}
\label{gmp}
G_k(q^2,t,t';\phi)=\raisebox{-10pt}{\includegraphics[width=60pt,keepaspectratio]{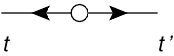}} \,,
\end{equation}
where an empty circle denotes $T \eta_k$. In the propagator the arrow indicates the direction of time: {\it i.e.}, in Eq.~(\ref{gp}) the function is nonzero only if $t>t'$, whereas there is no ordering in Eq.~(\ref{gmp}). 

In addition, two filled circles joined by a dashed line denote a vertex  obtained from the second cumulant $\mathsf{\Gamma}_{k2}$ and three filled circles joined by dashed lines a vertex obtained from the third cumulant $\mathsf{\Gamma}_{k3}$. There are no other vertices as we have set to zero the cumulants of order $p\geq 4$ in our ansatz for the effective average action. The legs of the vertices are associated with differentiation with respect to the fields. A leg with an incoming arrow indicates a derivative with respect to $\phi_{xt}$ and a leg with an outgoing arrow indicates a derivative with respect to $\hat\phi_{xt}$.

A number of diagrams in the above flows of the cumulants have been signaled by a (red) star. These diagrams need to be treated with special care. They indeed all contain a first derivative with respect to a physical field $\phi_a$ of a second or third random-force cumulant apparently evaluated when two replica fields (including $\phi_a$) are equal. As already stressed, the cumulants at $T=0$ develop linear cusps when two of their replica-field arguments become equal. One then expects that the derivatives in the direction of the cusp are singular when evaluated for exactly coinciding field arguments. This problem has already been encountered and solved in the perturbative FRG treatment of the REMM\cite{FRGledoussal-chauve} and, in our previous work,\cite{balog_eqnoneq} in the nonperturbative FRG treatment of the RFIM in and out of equilibrium. It can be worked out for the equilibrium and the depinning cases, but in two distinct and specific ways which we now illustrate. 

\subsection{Handling the cusp: An illustration}

\subsubsection{Depinning: The role of an infinitesimal velocity}

For illustration we consider the last diagram appearing in the flow of $\Delta_k$ in Eq.~(\ref{eq_2nd}) and marked by a red star. It involves a derivative of the third cumulant of the random force. There are four such diagrams having the same topology and we only consider the integral over $t_1,t_2$ of the one of the four (the integral over the internal momentum $q$ does not lead to any difficulty): 
\begin{eqnarray}
\label{ex_problematic}
I_{dep}(\phi_{1t_1},\phi_{2t_2})=\int_{t_3}\Bigg(\raisebox{-10pt}{\includegraphics[height=30pt,keepaspectratio]{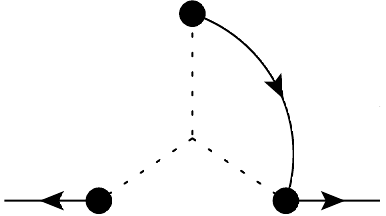}\put(-47,-8){\tiny $1,t_1$}\put(-19,-8){\tiny $2,t_2$}}\Bigg)=\int_{t_3}G_k^+(q^2;t_2-t_3)W_k^{(001)}(\phi_{1t_1},\phi_{2t_3},\phi_{2t_2}).
\end{eqnarray}
The response function introduces a time ordering, requiring that $t_{2}>t_3$. Because of the quasi-static driving, the dependence of the fields drifts infinitesimally slowly, $\phi_{a,t_a}\to\phi_a+v t_a$, and we need to evaluate the above expression in the limit $v\to 0^+$. Therefore, in the derivative of $W_k$ the difference of the last two arguments is $\phi_{2t_3}-\phi_{2t_2}\approx v(t_3-t_2) \to 0^-$. The ambiguity coming from the derivative in the presence of a cusp is then lifted because one does not consider the field difference to be $0$ but $0^-$, which chooses in a sense one side of the cusp and is unambiguous. By using the parametrization introduced in Sec.~\ref{approx} and the symmetries of the function $W_k$ discussed in this same section, one  can rewrite Eq.~(\ref{ex_problematic}) as
\begin{equation}
I_{dep}(y)=\int_{0}^{+\infty} dt' G_k^+(q^2;t')\left [\frac{W_k^{(01)}(0^-,\frac{2y}{\sqrt{3}})}{\sqrt{6}}-\frac{W_k^{(10)}(0^-,\frac{2y}{\sqrt{3}})}{\sqrt{2}}\right ].
\end{equation}  
After introducing the static propagator
\begin{equation}
P_k(q^2)=\frac 1{q^2+\widehat R_k(q^2)}
\end{equation}  
and using the symmetry of the derivatives, $W_k^{(01)}(-y,z)=W_k^{(01)}(y,z)$ and  $W_k^{(10)}(-y,z)=-W_k^{(10)}(y,z)$, we finally arrive at the expression
\begin{equation}
\label{ex_pr_3}
I_{dep}(y)=P_k(q^2)\left[\frac{W_k^{(01)}(0,\frac{2y}{\sqrt{3}})}{\sqrt{6}}+\frac{W_k^{(10)}(0^+,\frac{2y}{\sqrt{3}})}{\sqrt{2}}\right ]\,,
\end{equation} 
where $W_k^{(10)}(0^+,2y/\sqrt{3})$ is well defined in the presence of a cusp in the first argument.

\subsubsection{Equilibrium: The role of an infinitesimal temperature}

We now consider the same quantity in Eq~(\ref{ex_problematic}) as for the depinning case but for the equilibrium situation. This means in particular that the replica fields are independent of time. $I_{eq}(y)$ is then given by the same expression as in Eq~(\ref{ex_pr_3}) except that $0^+$ should be replaced by $0$. The result is therefore {\it a priori} ambiguous at zero temperature in the presence of a cusp. However, an infinitesimal temperature rounds the cusp and allows one to deal with an analytic theory. The equilibrium fixed point controlling the pinned phase is at zero temperature and the renormalized temperature is an irrelevant, albeit dangerously so, quantity. The rounding of the cusp is a nontrivial phenomenon and it takes place inside a thermal boundary layer that shrinks with the RG flow as one approaches the fixed point. The convergence to the zero-temperature fixed point is therefore nonuniform and requires some subtle treatment.\cite{balents-ledoussal,ledoussal10} However, in a nutshell, one can use the fact that in the presence of an infinitesimal temperature the cumulants are all well behaved within the boundary layer. This leads to symmetry properties that can be enforced first, before one takes the limit of zero temperature.

So, one may consider the expression  
\begin{equation}
\label{ex_pr_eq}
I_{eq}(y)=P_k(q^2)\left[\frac{W_k^{(01)}(0,\frac{2y}{\sqrt{3}})}{\sqrt{6}}+\frac{W_k^{(10)}(0,\frac{2y}{\sqrt{3}})}{\sqrt{2}}\right ]
\end{equation} 
in the presence of an infinitesimal temperature, use the symmetry properties characteristic of equilibrium in an analytic theory (within the thermal boundary layer) which entails that $W_k^{(10)}(y=0,z)=0$, and then take the limit of zero temperature. This leads to
\begin{equation}
\label{ex_pr_eq}
I_{eq}(y)=P_k(q^2)\,\frac{W_k^{(01)}(0,\frac{2y}{\sqrt{3}})}{\sqrt{6}}\,.
\end{equation} 
One can see that this expression is different than that of the depinning case. Infinitesimal velocity (for depinning) and infinitesimal temperature (for equilibrium) both lift the ambiguities associated with the presence of cusps in the functional dependence of the cumulants, but they do it differently and lead to different beta functions. This is exactly what was done before in the context of the perturbative FRG of the present model\cite{FRGchauve,FRGledoussal-chauve} and in our previous nonperturbative FRG investigation of the RFIM in and out of equilibrium.\cite{balog_eqnoneq}

\subsection{NP-FRG flow of the second and third cumulants: Explicit expressions}

By following the procedure outlined above, and after some tedious but rather systematic and straightforward manipulations, we arrive at explicit NP-FRG flow equations for $\Delta_k(y)$ and $W_k(y,z)$ both for equilibrium and for depinning at zero temperature. The equation for the second cumulant explicitly reads
\begin{eqnarray}
\label{flow_D_dep_1}
\partial_k\Delta_k(y)&=&-\left (\frac{\tilde{\partial}_k}{2}\int_q  P_k(q^2)^2 \right )\bigg (\Delta_k'(y)^2+[\Delta(y)-\Delta_k(0)]\Delta_k''(y)\bigg )\nonumber\\&
+& \left (\frac{\tilde{\partial}_k}{2}\int_q  P_k(q^2) \right )\bigg (\frac {\sqrt{6}}{6} \big [W_k^{(01)}(y,\frac{y}{\sqrt{3}})+\sqrt{3} W_k^{(10)}(y,\frac{y}{\sqrt{3}})-2 W_k^{(01)}(0,\frac{2y}{\sqrt{3}}) \big] \nonumber\\&
-&\lambda \frac{\sqrt 2}2 \big[-W_k^{(10)}(y,\frac{y}{\sqrt{3}})+\sqrt{3} W_k^{(01)}(y,\frac{y}{\sqrt{3}}) + 2W_k^{(10)}(0,\frac{2y}{\sqrt{3}})\big]\bigg)
\end{eqnarray} 
where $\lambda=0$ for equilibrium and $\lambda=1$ for depinning and a prime denotes a derivative with respect to the argument. The variable $y$ is restricted $y\geq 0$ and the arguments of $W_k(y,z)$ are restricted to the upper right (shaded) triangle in Fig.~\ref{fig_w_dep}, {\it i.e.}, to $y\geq 0,\, z\geq y/\sqrt 3$ [so, for instance, $W_k^{(10)}(0,\frac{2y}{\sqrt{3}})$ should be interpreted as $W_k^{(10)}(0^+,\frac{2y}{\sqrt{3}})$ and $W_k^{(01)}(y,\frac{y}{\sqrt{3}})$ as $W_k^{(01)}(y,\frac{y}{\sqrt{3}}+0^+)$, with $y\geq 0$]. 

The above flow equation can be rewritten in a more compact form as
\begin{equation}
\begin{aligned}
\label{flow_D_dep_2}
&\partial_k\Delta_k(y)=\\&- \frac 12 \left (\frac{\tilde{\partial}_k}{2}\int_q  P_k(q^2)^2 \right )\partial_y^2\big ([\Delta(y)-\Delta_k(0)]^2\big )+
\frac {\sqrt 2}{2} \left (\frac{\tilde{\partial}_k}{2}\int_q  P_k(q^2) \right )\partial_y \big [W_k(y,\frac{y}{\sqrt{3}})-W_k(0,\frac{2y}{\sqrt{3}})\big] 
\\&-\lambda \frac{\sqrt 2}2 \left (\frac{\tilde{\partial}_k}{2}\int_q  P_k(q^2)\right )\big[-W_k^{(10)}(y,\frac{y}{\sqrt{3}})+\sqrt{3} W_k^{(01)}(y,\frac{y}{\sqrt{3}}) 
+ 2 W_k^{(10)}(0,\frac{2y}{\sqrt{3}})\big]\,.
\end{aligned}
\end{equation} 
From the above expression one immediately sees that in the equilibrium case ($\lambda=0$) the beta function for $\Delta_k(y)$ is a total derivative with respect to $y$. Provided the cumulants are well enough behaved for large arguments, this implies that the integral $\int_0^{+\infty} dy \Delta_k(y)$ (or $\int_0^{\infty} dy \Delta_k(y)$ for a periodic disorder of period $1$) does not flow and remains equal to its bare value (see also below). This is related to the ``potentiality property'',\cite{FRGledoussal-chauve} according to which the renormalized random force is the derivative of a renormalized random potential and, as a result, the second cumulant $\Delta_k(y)$ is (up to a sign) the second derivative of the second cumulant of this random potential. It is easy to check that the above flow equation for $\Delta_k(y)$ is indeed the second derivative of the exact flow equation for the second cumulant of the random potential obtained in Ref. [\onlinecite{ledoussal10}], with the additional approximation that the cumulants are purely local.

To search for the fixed points controlling the physics at long distance and long times one must introduce scaling dimensions and cast the flow equations such as Eq.~(\ref{flow_D_dep_2}) in a dimensionless form. As one knows that the fixed points are zero-temperature ones both for equilibrium and depinning, a properly defined renormalized temperature $T_k$ must be defined, {\it e.g.}, by comparing the second cumulant and the first one, $T_k \propto k^2/\Delta_k$ where $\Delta_k$ characterizes the amplitude of the second cumulant of the renormalized random force and can be for instance chosen as its value at the origin $y=0$. As explained in our previous work on the RFIM in and out of equilibrium,\cite{balog_eqnoneq} this temperature coincides with the temperature entering via the thermal noise in the original Langevin equation when the system is at equilibrium but it does not when the system is athermally driven. In the latter case the renormalized ``temperature'' is unrelated to a thermal bath and just describes the relative scaling of successive disorder cumulants.

By using the following scaling dimensions,
\begin{equation}
\begin{aligned}
\label{eq_scaling_dim}
y,\, z \sim k^{-\zeta}\,,\; T_k \sim k^\theta\,, \; \Delta_k \sim k^{4-d-2\zeta}\,, \; W_k\sim k^{6-2d-3\zeta}\,,\; \widehat R_k \sim k^2
\end{aligned}
\end{equation}
with $\theta=d-2+2\zeta$, one can recast Eq.~(\ref{flow_D_dep_2}) in a dimensionless form,
\begin{equation}
\begin{aligned}
\label{flow_D_dep_3}
\partial_s\delta_k(y)= &(d-4+2\zeta)\delta_k(y) -\zeta\, y \delta_k'(y)+ \frac 12 I_3\, \partial_y^2\big ([\delta_k(y)-\delta_k(0)]^2\big )\\&
- \frac {\sqrt 2}{4}I_2\, \partial_y \big [w_k(y,\frac{y}{\sqrt{3}})-w_k(0,\frac{2y}{\sqrt{3}})\big] 
\\& + \lambda \frac {\sqrt 2}{4}I_2\, \big[-w_k^{(10)}(y,\frac{y}{\sqrt{3}})+\sqrt{3} w_k^{(01)}(y,\frac{y}{\sqrt{3}}) +2 w_k^{(10)}(0,\frac{2y}{\sqrt{3}})\big]\,,
\end{aligned}
\end{equation} 
where we have introduced the dimensionless RG ``time'' $s=\ln (k/\Lambda)$ and the dimensionless quantities $I_p=-k^{-d+2(p-1)}\tilde{\partial}_s\int_q  P_k(q^2)^{p-1}/(p-1)=\int_{\hat q}  \partial_s \hat r_k(\hat q^2) (\hat q^2+\hat r_k(\hat q^2))^{-p}$, with $\hat q=q/k$ and $\hat r(\hat q^2)=\widehat R(q^2)/k^2$; we have used lower case letters for the dimensionless cumulants (but kept for convenience the same notation for the dimensionless and dimensionful fields, hoping that this will not cause any confusion). The arguments of the functions are restricted as explained below Eq.~(\ref{flow_D_dep_1}).

With more effort but along the same lines one can derive the flow equation for the third cumulant, which in a dimensionless form reads
\begin{equation}
  \label{flow_w_dep}
\partial_s w_k(y,z)= (2d-6+3\zeta) w_k(y,z) -\zeta(y\partial_y+z\partial_z) w_k(y,z) +\beta_{w,k}^{(eq)}(y,z)+\lambda\, \Delta\beta_{w,k}(y,z),
\end{equation}
where, again, $\lambda=0$ for equilibrium and $\lambda=1$ for depinning. The beta function for equilibrium $\beta_{w,k}^{(eq)}(y,z)$ is given in Appendix \ref{app_w_beta} and $\Delta\beta_{w,k}(y,z)$ is equal to
\begin{equation}
\begin{aligned}
\label{flow_w_dep-eq}
&\Delta\beta_{w,k}(y,z)=\\&
 I_3 \Big\{\frac {\sqrt{3}}6\delta_k '(\frac{y}{2}+\frac{\sqrt{3}z}2) \big [w_k^{(01)}(0,\frac{2 y}{\sqrt{3}})-\sqrt 3 w_k^{(10)}(0,\frac{2 y}{\sqrt{3}})
 +2 w_k^{(01)}(-\frac{y}{2}+\frac{\sqrt{3} z}2,-\frac{\sqrt{3} y}{6} +\frac z2 )\big ]
\\&-\frac {\sqrt{3}}3\delta_k '(-\frac{y}{2}+\frac{\sqrt{3} z}2) \big [ w_k^{(01)}(y,\frac{y}{\sqrt{3}})- w_k^{(01)}(\frac{y}{2}+\frac {\sqrt{3} z}2,\frac{\sqrt{3} y}{6} + \frac z2)\big ]
\\&+\delta_k '(y) \big [w_k^{(10)}(0,-\frac{y}{\sqrt{3}}+z)- w_k^{(10)}(0,\frac{y}{\sqrt{3}}+z)\big ]\Big\}
\\&+\frac {3\sqrt 2}2 I_4 \delta_k '(0) \Big \{\delta_k '(y)\big [\delta_k '(\frac{y}{2}+\frac{\sqrt{3} z}2)- \delta_k '(-\frac{y}{2} +\frac{\sqrt{3} z}2)\big ]
+\delta_k '(-\frac{y}{2}+\frac{\sqrt{3} z}2) \delta_k '(\frac{y}{2}+\frac{\sqrt{3} z}2)\Big\},
\end{aligned}
\end{equation} 
where the arguments of the functions are restricted as explained below Eq.~(\ref{flow_D_dep_1}).

Before discussing the solution of these NP-FRG equations we move on to a derivation of the RG flow for the friction.

\subsection{Renormalization of the friction}
\label{sec_friction}

We briefly comment on the RG flow of the friction. With the nonperturbative ansatz and the RG prescription given in Sec.~\ref{approx}, it is obtained from 
\begin{equation}
\label{fric1}
  \partial_k \frac{\delta^2\mathsf{\Gamma}_{k1}[\Phi]}{\delta\phi_{xt}\delta\hat{\phi}_{xt'}}\Big\vert_{unif}= -\frac{\tilde{\partial}_k}{2}\int_q\int_{t_i}\Bigg(\raisebox{-15pt}{\includegraphics[height=30pt,keepaspectratio]{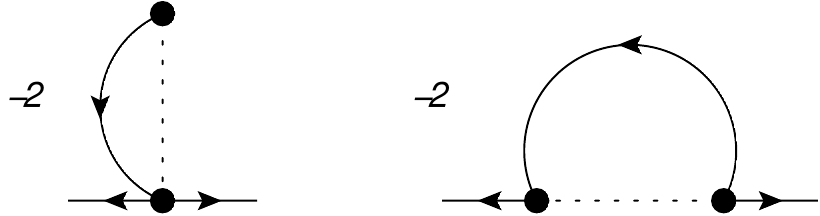}}\Bigg)\,,
\end{equation}
which again involves derivatives of the second cumulant $\Delta_k$ evaluated for the same replica.

In the quasi-statically driven case, by using the same procedure as before in which the cusp can be regularized by an infinitesimal velocity, we arrive at
\begin{equation}
  \label{fric1_fin}
  \partial_k\eta_k=-\left (\frac {\tilde{\partial}_k}{2}\int_q P_k(q^2)^2\right )\Delta_k''(0^+) \,\eta_k\,,
\end{equation}
where the quantity $\Delta_k''(0^+)$ is unambiguous even in the presence of a cusp. In a dimensionless form, the above equation simply reads $\partial_s \ln\eta_k=I_3\delta_k''(0^+)$, which via the scaling dimension of the friction, $\partial_s \ln\eta_k=2-z$, allows us to derive the dynamical exponent $z$ [not to be confused with the field $z$ introduced in Eq.~(\ref{eq_field_parametrization})].

The relaxation to equilibrium is more involved as one needs to keep a small temperature $T$ (otherwise there is no relaxation at all). The time-reversal symmetry and fluctuation-dissipation theorem are now satisfied but, as has been described and proven in great detail before,\cite{balents-doussal_dyn,balog_activated} the critical slowing down is now of an activated form: The relaxation time $\tau_k=\eta_k/k^2$ does not go as $k^{-z}$ as usual, but it is rather its logarithm $\ln \tau_k$ that grows as a power law $k^{-\psi}$ as $k\to 0$.\cite{fisher_activated} It can be seen from Eq.~(\ref{fric1_fin}) that in the absence of regularization by an infinitesimal velocity $\Delta_k''(0^+)$ should be naively replaced by $\Delta_k''(0)$, which then diverges in the presence of a cusp. The way out of this is to consider a nonzero temperature $T$, which under renormalization and considered near the fixed point becomes as small as wanted since $T_k\sim k^\theta T$ with $\theta>0$, so that one can use the property that the cusp is rounded in a thermal boundary layer.\cite{chauve_creep,balents-doussal_dyn,ledoussal10} This directly leads to an activated dynamical scaling with the exponent $\psi=\theta$.\cite{chauve_creep,balents-doussal_dyn} (Note that the scenario is similar but more complicated in the case of the RFIM as the exponent $\psi$ is equal to the temperature exponent $\theta$ for $d<d_{DR}\approx 5.1$ but decreases continuously for larger $d$ to reach $0$ at the upper critical dimension $d_{uc}=6$.\cite{balog_activated})

\section{Results}
\label{sec_results}

\subsection{Relation with the perturbative FRG in the $\epsilon=4-d$ expansion}

One easily checks that near $d=4$ and at any of the relevant fixed points the second cumulant of the random force is of order $\epsilon$ whereas the third one is of order $\epsilon^3$, with $\epsilon=4-d$. At leading order in $\epsilon$  the above NP-FRG equations for $\delta_k(y)$, Eq.~(\ref{flow_D_dep_2}), then reduces to the exact one-loop expression first derived in Refs.~[\onlinecite{fisher86-pin,nattermann_dep,narayan92}], expression which is the same for equilibrium and for depinning. The same is true for the flow equation for the friction.

One can proceed further and compare the expansion of our NP-FRG equations at the next order in $\epsilon$ with the exact results obtained at 2-loop order for equilibrium and depinning by Chauve et al.\cite{FRGchauve,FRGledoussal-chauve} To do so we first solve the fixed-point equation for the third cumulant $w_k(y,z)$ at order $\epsilon^3$, which only requires the knowledge of the function $\delta_k(y)$ and we insert the result (which is given in Appendix \ref{app-epsilon_exp}) in the RG flow equation for $\delta_k(y)$. This straightforwardly leads to 
\begin{eqnarray}
\label{eps2_eq_dep}
\partial_s \delta_k(y)&= (d-4+2\zeta)\delta_k(y)-\zeta y \delta_k'(y)+\frac{1}{2}I_3 \big([\delta_k(y)-\delta_k(0)]^2\big)''\nonumber\\
&-\frac{3}{4} I_2 I_4 \big(\delta_k'(y)^2 [\delta_k(y)-\delta_k(0)]{\color{red}\pm}\delta_k(y)\delta'(0^+)^2\big)''+{\rm O}(\epsilon^4).
\end{eqnarray}
where the ``$+$" sign in the last term is for depinning and ``$-$" sign for equilibrium and the $I_p$'s have been defined below Eq.~(\ref{flow_D_dep_3}). One can rescale the field $y$ to absorb the factor $I_3$ and the above expression in then in a form similar to the exact 2-loop result in Refs.~[\onlinecite{FRGchauve,FRGledoussal-chauve}], except that the prefactor of the last term is $3 I_2 I_4/(4I_3^2)$ instead of the exact value of $1/2$.

As anticipated, our NP-FRG ansatz is not exact at the 2-loop order (but it accounts for all loop contributions in an approximate way, and its numerical predictions will be tested in the following section). The coefficient $3 I_2 I_4/(4I_3^2)$ has some weak dependence on the choice of the IR cutoff function and is found around $0.75\pm 0.05$, to be compared with the exact value of $0.5$.

\subsection{Numerical results for the equilibrium and depinning fixed points}

By numerically solving the equations for the second and third cumulants one can find fixed points for any arbitrary dimension $d$ (we actually studied for $0\leq d\leq 4$) and for both equilibrium and depinning. Our goal is not to dwell on the physical interpretation of the results, since this has been extensively done by several  authors using the perturbative FRG.\cite{fisher86-pin,nattermann_dep,narayan92,FRGchauve,FRGledoussal-chauve,doussal-wiese-review,husemann18} We rather assess the accuracy of our findings when compared to these previous studies and to computer simulations.

For the equilibrium situation, we have considered the cases of a random-bond disorder and a random-field disorder. (This could be easily extended to the class of periodic disorder corresponding to pinned charge density waves but we do not think that it will bring any new insight for our purpose.)  In the random-field case, one finds as in previous studies\cite{grinstein-ma82,fisher86-pin,FRGledoussal-chauve,husemann18} that the large scale behavior is characterized by a roughness exponent exactly given by $\zeta=(4-d)/3$. This follows directly from integrating Eq.~(\ref{flow_D_dep_3}) with $\lambda=0$ between $0$ and $\infty$ with minor assumptions on the boundary conditions [namely, $y\delta_k(y)\to 0$ and $w_k(y,y/\sqrt 3)=w_k(0,2y/\sqrt 3)\to 0$ when $y\to \infty$], which leads to $0=(4-d-3\zeta)\int_0^{\infty}dy\delta_*(y)$  at the fixed point. For the random-field disorder $\int_0^{\infty} dy \delta_k(y)>0$ all along the flow, which gives the announced result. 

\begin{figure}[h!]
  \begin{center}
 \includegraphics[width=0.8\linewidth]{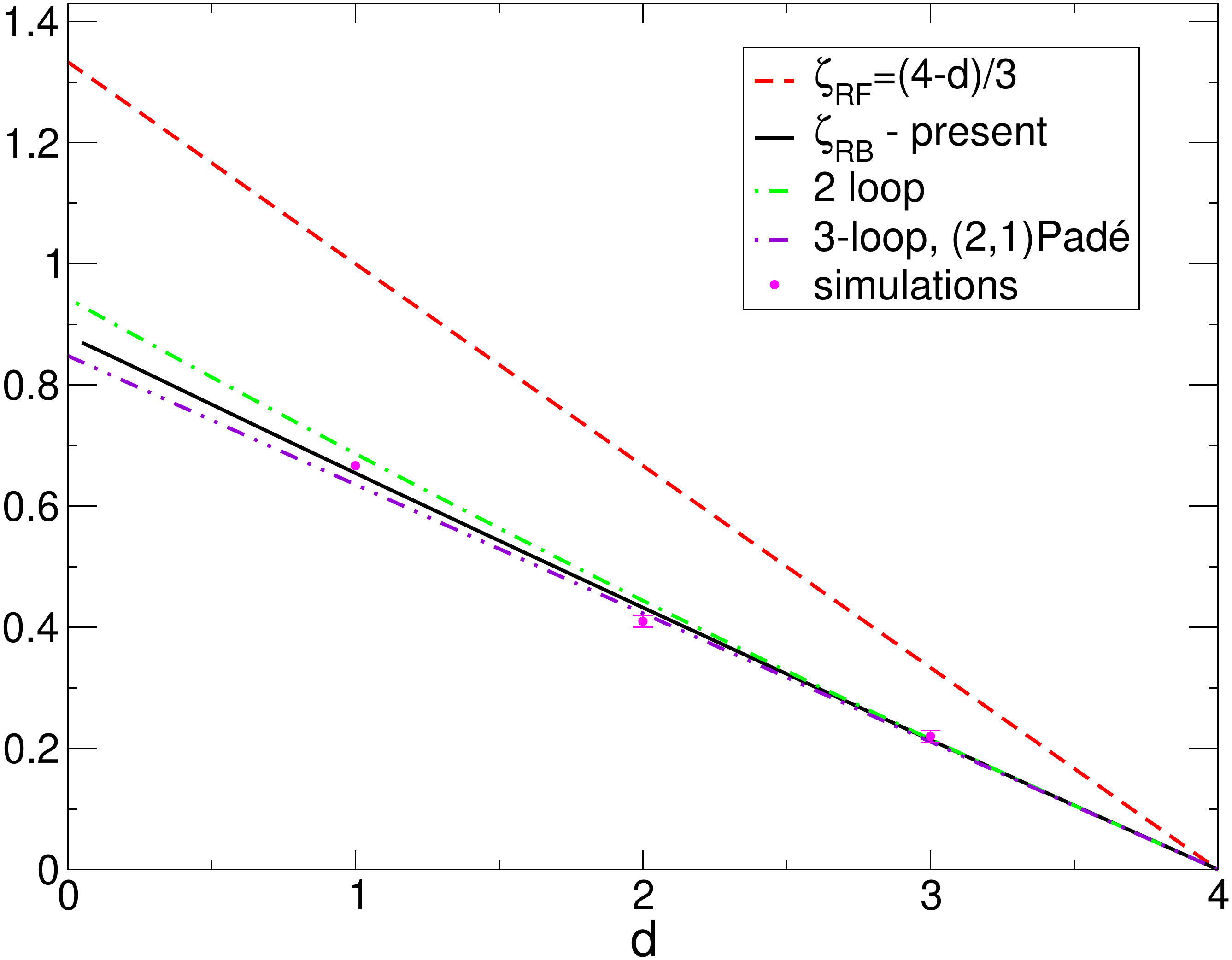}
    \caption{Roughness exponent $\zeta$ characterizing the pinned manifold phase at equilibrium as a function of dimension $d$ for random-field disorder [then $\zeta=(4-d)/3$, upper (red) full line] and random-bond disorder. For the latter, the (black) full line is the present NP-FRG prediction, the (green) dashed-dotted line is the 2-loop prediction,\cite{FRGledoussal-chauve} the dotted line the $(2,1)$-Pad\'e resummation of the 3-loop result,\cite{husemann18} and the symbols are simulation results.\cite{delta_eq,zeta_eq_RB}}\label{Fig_zeta_equil}
  \end{center}
\end{figure}

Random-bond disorder on the other hand is characterized by $\int_0^{\infty} dy \delta_k(y)=0$, which therefore does not lead to any constraint on the value of $\zeta$. The fixed point for random-bond disorder has then to be found numerically by solving the two coupled equations for $\delta_*(y)$ and $w_*(y,z)$. We plot in Fig.~\ref{Fig_zeta_equil} the resulting exponent $\zeta$ as a function of $d$. The agreement with computer simulation results and state-of-the-art perturbative FRG predictions is excellent. Note that compared to the 3-loop perturbative FRG result in powers of $\epsilon=4-d$,\cite{husemann18} there is no need to proceed to a Pad\'e resummation to obtain good results in $d=1$ and $d=2$. This is a strength of the NP-FRG approach which gives an accurate description in all dimensions. We provide  some details on the numerical resolution and the choice of IR cutoff function in Appendix~\ref{app_numerics}.

The relevant fixed points that describe the large-scale behavior of the pinned manifold in equilibrium are fully attractive. In the case of the random-bond disorder we have also numerically found a series of additional fixed points with an increasing number of relevant directions. The physical interpretation of these fixed points is unclear but the finding illustrates the accuracy of the present method. This is discussed in more detail in Appendix~\ref{app_multicriticalFP}, where we also give more results concerning the spectrum of eigenvalues around the attractive fixed point.

\begin{figure}[h!]
  \begin{center}
     \includegraphics[width=0.8\linewidth]{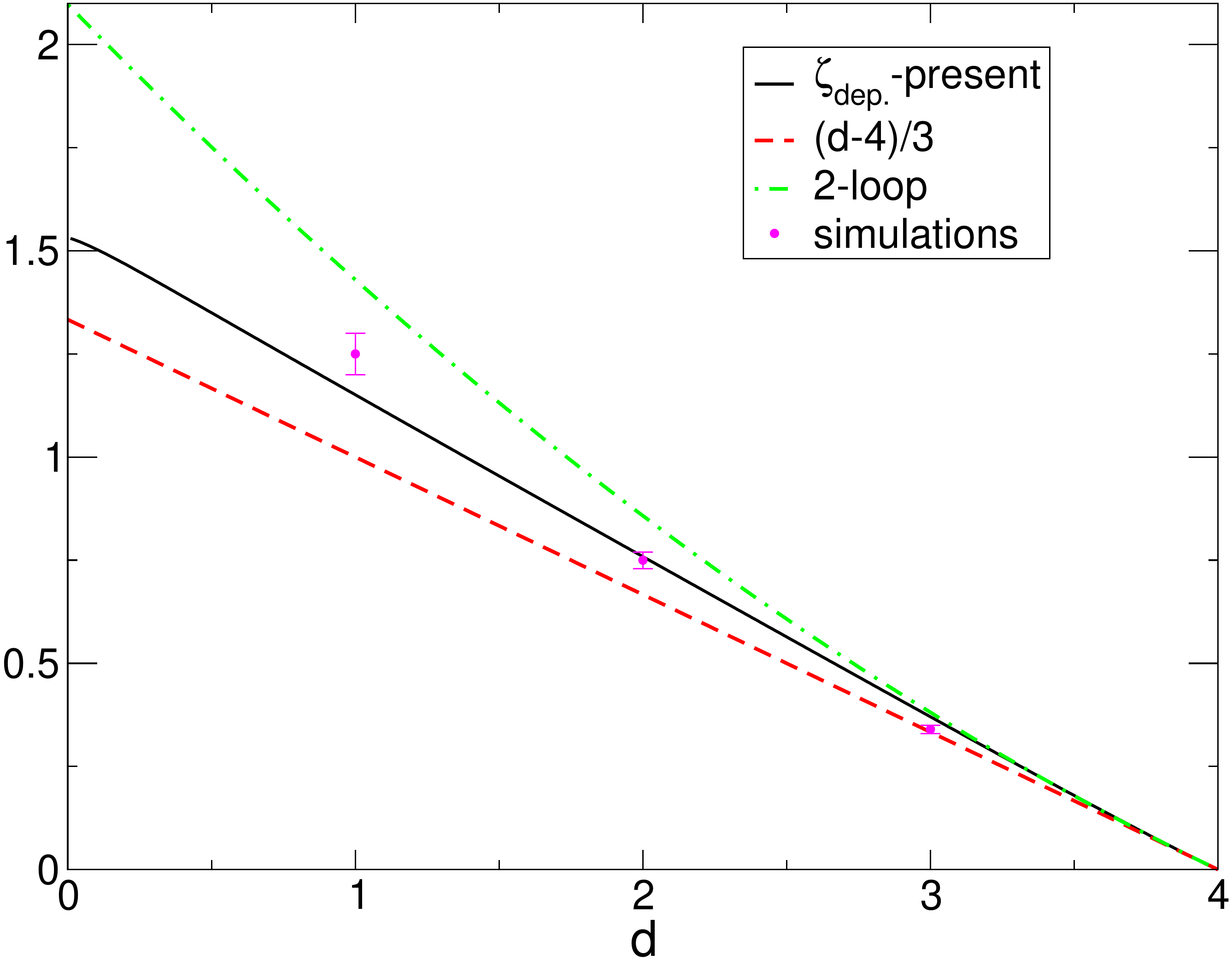}
    \caption{Roughness exponent $\zeta$ at the depinning transition as a function of dimension $d$ for nonperiodic disorder. The (black) full line is the present NP-FRG prediction, the (green) dashed-dotted line is the 2-loop prediction,\cite{FRGledoussal-chauve} and the symbols are simulation results.\cite{zeta_dep-1D,zeta_dep-2-3D} The (red) dashed line is the equilibrium value for random-field disorder.} \label{Fig_zeta_depin}
  \end{center}
\end{figure}

We now discuss the athermal quasi-static driven case at the depinning transition. As known from previous studies,\cite{narayan92,FRGledoussal-chauve,chauve_creep} one finds no fixed point associated with random-bond disorder. Whether starting from initial conditions with random-bond or random-field disorder, we indeed obtain that the RG flow goes to a random-field disorder fixed point and that this fixed point is different than the equilibrium one. This is an important output that our nonperturbative but approximate FRG approach correctly captures. (There is also a fixed point describing depinning in the presence of with periodic disorder but we have not studied it.) In Fig.~\ref{Fig_zeta_depin} we plot the roughness exponent $\zeta$ as a function of dimension and compare our results to those obtained from the 2-loop FRG and computer simulations. Here too the agreement is excellent.

At the depinning threshold,  scaling involves an additional exponent on top of the roughness one, {\it e.g.}, the dynamical exponent $z$ obtained from the flow of the friction: see Sec.~\ref{sec_friction}. (All the other exponents can be obtained from $\zeta$ and $z$.\cite{fisher_dep,nattermann_dep,narayan92}) The output of our NP-FRG approach is $z=1.69$, $1.33$, and $0.97$ in $d=3$, $2$, and $1$, respectively, to be compared with the 2-loop predictions, $z=1.73$, $1.38$, and $0.94$.\cite{FRGledoussal-chauve}

\begin{figure}[h!]
  \begin{center}
   \includegraphics[width=0.8\linewidth]{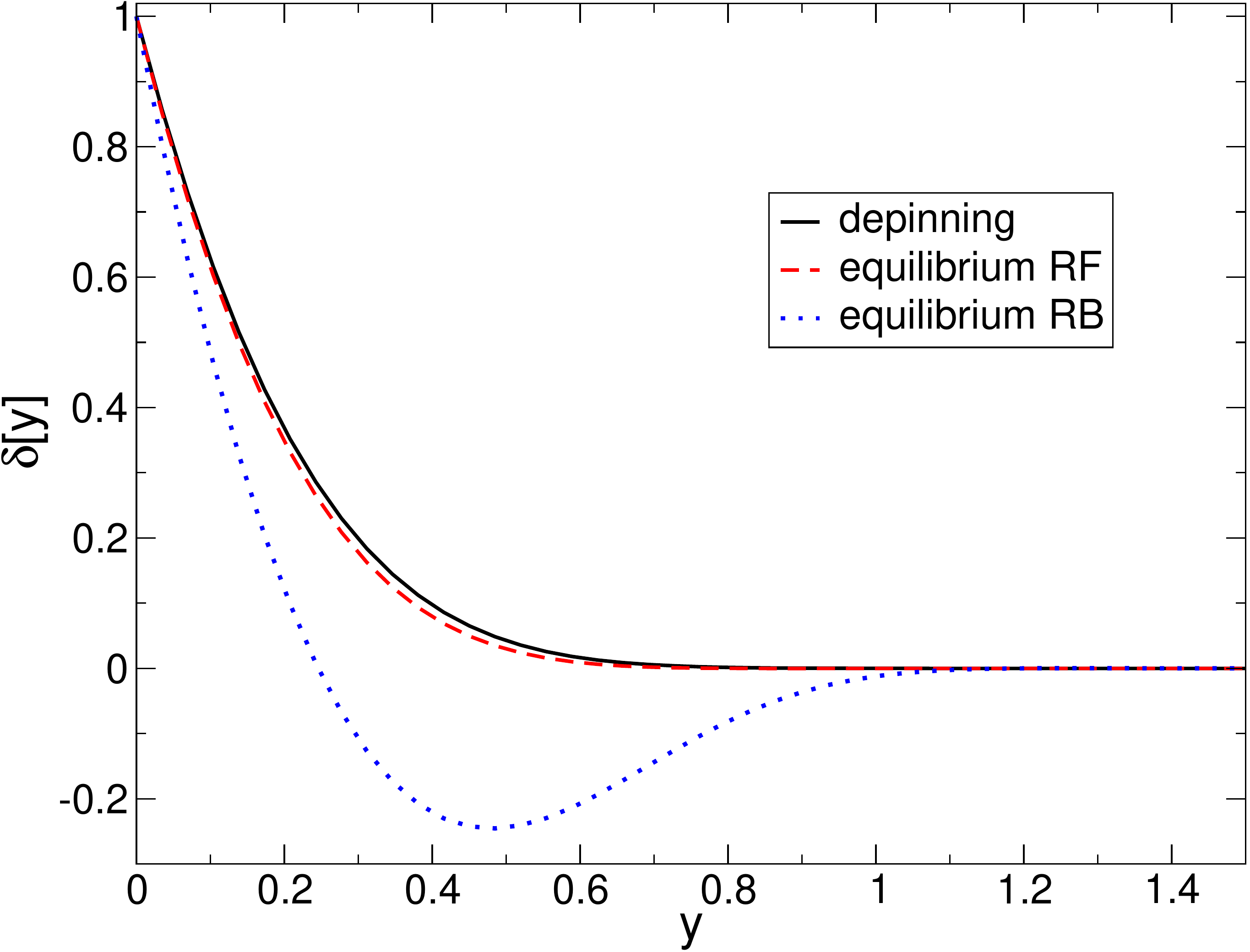}
    \caption{Second cumulant of the random force $\delta_*(y)$ at the fixed point for equilibrium and depinning in $d=2$ ({\it i.e.}, an interface in a $3$-$d$ disordered medium): depinning (full line), random-bond disorder at equilibrium (dotted, lower, curve), random-field disorder at equilibrium (dashed, upper, curve). At the depinning threshold the functions for random-field and random-bond disorder exactly coincides. The theoretical predictions compare very well with the 2-loop perturbative FRG\cite{FRGledoussal-chauve} and the simulation results.\cite{delta_eq,delta_dep}}\label{Fig_delta(y)}
  \end{center}
\end{figure}

For completeness we also display in Fig.~\ref{Fig_delta(y)} the second cumulant of the random force at the fixed point, $\delta_*(y)$, for both equilibrium and depinning and nonperiodic disorder in $d=2$. The function, which displays a cusp around $y=0$, has been computed via the 2-loop perturbative FRG\cite{FRGledoussal-chauve} and measured in computer simulations as well.\cite{delta_eq,delta_dep}  Its shape does not vary much with the dimension $d$, as we also find (compare, for {\it e.g.}, the equilibrium random-bond fixed-point function for $d=3$ in Fig.~\ref{fig_RBFPS_d3} of Appendix~\ref{app_multicriticalFP} and for $d=2$ in Fig.~\ref{Fig_delta(y)}). Our theoretical predictions compare very well with the 2-loop and simulation  results.

\section{Conclusion}

We have applied the nonperturbative FRG formalism previously developed by us to investigate the critical behavior of the RFIM in and out of equilibrium\cite{balog_eqnoneq} to the well-studied case of an elastic manifold in a random environment. By following the same procedure as used before for the RFIM we have recovered the main long-distance and long-time properties of the random elastic manifold model, both in the pinned phase at equilibrium and at the depinning threshold in the athermally and quasi-statically driven case. Our method predicts critical exponents and fixed-point functions that are in excellent agreement with the best known results. As a side comment we note that, although there is no requirement for a nonperturbative treatment {\it per se} in the present problem of the random elastic manifold (at odds with the RFIM), our method that is based on a truncation of the cumulant expansion and of the derivative expansion of the effective action at fixed dimension $d$ provides critical exponents and fixed-point functions in all dimensions $0<d\leq 4$ with no need to invoke resummation techniques (contrary to the perturbative FRG based on an expansion in $\epsilon=4-d$). In any case, this successful benchmarking of our theoretical approach gives strong support to the results that we have previously obtained for the RFIM, in particular concerning the distinct universality classes of the equilibrium and out-of-equilibrium (hysteresis) critical points below a critical dimension $d_{DR}\approx 5.1$.

\begin{acknowledgments}
We thank K.J. Wiese for fruitful discussions. IB acknowledges the support of the Croatian Science Foundation Project No. IP-2016-6-7258 and the QuantiXLie Centre of Excellence, a project cofinanced by the Croatian Government and European Union through the European Regional Development Fund - the Competitiveness and Cohesion Operational Programme (Grant KK.01.1.1.01.0004). IB also thanks the LPTMC for its hospitality and the CNRS for funding during the spring of 2019.  
\end{acknowledgments}

\appendix

\section{Expressions for the beta function of the third cumulant common to equilibrium and depinning}
 \label{app_w_beta}  

By following the procedure outlined in the main text, we have derived from Eq.~(\ref{eq_3rd}) the NP-FRG flow equations for the third cumulant $w_k(y,z)$ for the equilibrium and the depinning cases. Below we give the explicit expression for the contributions to the nontrivial part of the beta functions that is common to equilibrium and depinning: 
\begin{equation}
\begin{aligned}
\label{flow_w_noproblem}
&\beta_{w,k}^{(eq)}=\frac {3 \sqrt{2}}4 I_4\Big\{-\delta_k'(y)^2 \big [\delta_k'(\frac{1}{2} (-y+\sqrt{3} z))+\delta_k'(\frac{1}{2} (y+\sqrt{3} z))\big ] 
\\&
+\delta_k '(\frac{1}{2} (-y+\sqrt{3} z))^2 \delta_k '(\frac{1}{2} (y+\sqrt{3} z))+\delta_k '(y) \Big (\delta_k ''(\frac{1}{2} (-y+\sqrt{3} z))\big [-\delta_k (0)+\delta_k (y) 
\\&
+\delta_k (\frac{1}{2} (-y+\sqrt{3} z))-\delta_k (\frac{1}{2} (y+\sqrt{3} z))\big ]+ \delta_k ''(\frac{1}{2} (y+\sqrt{3} z)) \big [\delta_k (0)-\delta_k (y) 
\\&
+\delta_k (\frac{1}{2} (-y+\sqrt{3} z))-\delta_k (\frac{1}{2} (y+\sqrt{3} z))\big ] 
+\big [\delta_k '(\frac{1}{2} (-y+\sqrt{3} z))^2-\delta_k '(\frac{1}{2} (y+\sqrt{3} z))^2\big ] \Big )
\\&
+\delta_k '(\frac{1}{2} (-y+\sqrt{3} z)) \Big (\delta_k ''(y) \big [\delta_k (0)-\delta_k (y) -\delta_k (\frac{1}{2} (-y+\sqrt{3} z))+\delta_k (\frac{1}{2} (y+\sqrt{3} z))\big ]
\\&
+ \delta_k ''(\frac{1}{2} (y+\sqrt{3} z))\big [-\delta_k (0)-\delta_k (y)+\delta_k (\frac{1}{2} (-y+\sqrt{3} z)) +\delta_k (\frac{1}{2} (y+\sqrt{3} z))\big ] +\delta_k '(\frac{1}{2} (y+\sqrt{3} z))^2 \Big )
\\&
+\delta_k '(\frac{1}{2} (y+\sqrt{3} z)) \Big (\delta_k ''(y) \big[\delta_k (0)-\delta_k (y)+\delta_k (\frac{1}{2} (-y+\sqrt{3} z))-\delta_k (\frac{1}{2} (y+\sqrt{3} z))\big ]
\\&
+  \delta_k ''(\frac{1}{2} (-y+\sqrt{3} z)) \big [-\delta_k (0)-\delta_k (y)+\delta_k (\frac{1}{2} (-y+\sqrt{3} z))+\delta_k (\frac{1}{2} (y+\sqrt{3} z))\big ]\Big )\Big\}
\\&
+\frac{I_3}{6} \Big\{3 \big [w_k(0,-\frac{y}{\sqrt{3}}+z) +w_k(0,\frac{y}{\sqrt{3}}+z) -2 w_k(y,z) \big ] \delta_k ''(y)
\\&
+3\big [w_k(y,\frac{y}{\sqrt{3}})-2 w_k(y,z)+w_k(\frac{1}{2} (y+\sqrt{3} z),\frac{1}{6} (\sqrt{3} y+3 z)) \big ]\delta_k ''(\frac{1}{2} (-y+\sqrt{3} z))
\\&+
3\big [w_k(0,\frac{2 y}{\sqrt{3}})-2 w_k(y,z) +w_k(\frac{1}{2} (-y+\sqrt{3} z),\frac{1}{6} (-\sqrt{3} y+3 z))\big ] \delta_k ''(\frac{1}{2} (y+\sqrt{3} z))
\\&
+2 \big [3 \delta_k (0)+\delta_k (y)-2 \delta_k (\frac{1}{2} (-y+\sqrt{3} z))-2 \delta_k (\frac{1}{2} (y+\sqrt{3} z))\big ] w_k^{(02)}(y,z)
\\&
+4 \sqrt{3} \big [\delta_k (\frac{1}{2} (-y+\sqrt{3} z))-\delta_k (\frac{1}{2} (y+\sqrt{3} z))\big ] w_k^{(11)}(y,z)
+6 \big [\delta_k (0)-\delta_k (y)\big ] w_k^{(2,0)}(y,z)
\\&
+3\delta_k '(\frac{1}{2} (y+\sqrt{3} z)) \big [-2 \sqrt{3} w_k^{(01)}(y,z) -2 w_k^{(10)}(y,z) + \frac{2 \sqrt{3}}3 w_k^{(01)}(0,\frac{2 y}{\sqrt{3}})
\\&
+\frac{\sqrt{3}}3 w_k^{(01)}(\frac{1}{2} (-y+\sqrt{3} z),\frac{1}{6} (-\sqrt{3} y+3 z))+ w_k^{(10)}(\frac{1}{2} (-y+\sqrt{3} z),\frac{1}{6} (-\sqrt{3} y+3 z)) \big ]
\\&
+3\delta_k '(\frac{1}{2} (-y+\sqrt{3} z)) \big [-2 \sqrt{3} w_k^{(01)}(y,z)+2 w_k^{(10)}(y,z)-\frac{\sqrt{3}}3 w_k^{(01)}(y,\frac{y}{\sqrt{3}})
\\&
+\frac{\sqrt{3}}3 w_k^{(01)}(\frac{1}{2} (y+\sqrt{3} z),\frac{1}{6} (\sqrt{3} y+3 z))
- w_k^{(10)}(y,\frac{y}{\sqrt{3}})+ w_k^{(1,0)}(\frac{1}{2} (y+\sqrt{3} z),\frac{1}{6} (\sqrt{3} y+3 z))\big ]
\\&
+2\sqrt{3} \delta_k '(y) \big [- w_k^{(01)}(0,-\frac{y}{\sqrt{3}}+z)+ w_k^{(01)}(0,\frac{y}{\sqrt{3}}+z)-2\sqrt{3} w_k^{(10)}(y,z)\big ]\Big\}\,,
\end{aligned}
\end{equation}
where we recall that we have introduced the dimensionless quantities $I_p=-k^{-d+2(p-1)}\tilde{\partial}_s\int_q  P_k(q^2)^{p-1}/(p-1)=\int_{\hat q}  \partial_s \hat r_k(\hat q^2) (\hat q^2+\hat r_k(\hat q^2))^{-p}$, with $\hat q=q/k$ and the dimensionless IR cutoff function $\hat r(\hat q^2)=\widehat R(q^2)/k^2$. The arguments $y$ and $z$ are restricted to $y\geq 0$ and $z\geq y/\sqrt 3$, see Sec.~\ref{approx}.

\section{Numerical resolution and choice of the IR cutoff function}
\label{app_numerics}

The numerical solution of the flow equations is obtained by custom built programs in Fortran 90. The details of implementation are essentially the same as those for the RFIM case which are discussed in Appendix B of Ref.~[\onlinecite{balog_eqnoneq}] and will not be repeated here. The only notable differences are the following. First, the present problem is discretized by using a triangular lattice (within the triangular domain indicated in Fig.~\ref{fig_w_dep}) for the third cumulant $w_k(y,z)$  and a finite line segment for the second cumulant $\delta_k(y)$. We found that accurate and robust results are achieved by choosing the following numerical parameters for the discretization of the line segment: $dy=\frac{2}{\sqrt{3}}dz$ with $dz=0.04$ and $n=101$ points. With these parameters, the field $y$ extends from $0$ to $3.464$. Such a choice then implies a range of the field $z$ between $0$ and $4$ and a total of $5061$ points on the triangular grid. Secondly, in order to obtain stable flows (and the fixed-point determination by the adapted Newton-Raphson method as well), we used a boundary condition which amounts to setting the functions to $0$ both on the outer edge of the triangle for the third cumulant and at the ultimate point of the segment and beyond for the second cumulant. We checked that with the parameters $n$ and $dz$ that we use, the functions would anyhow be extremely small at the boundary ($\approx 10^{-9}$), were they not forced to be exactly $0$. The procedure introduces small spurious oscillations near the boundary but it does not affect the functions in most of the grid (especially in the relevant region of small to moderate fields) and does not affect the critical exponents either.

\begin{table}
\begin{center}
  \begin{tabular}{ c | c | c }
     d & $\zeta_{dep}$ & $a_{PMS}$\\ \hline\hline
     $3.9$ & $0.03403$ & $7.15$ \\ \hline
     $3$   & $0.3703$ & $4.80$ \\ \hline
     $2$   & $0.7581$ & $3.20$ \\ \hline
     $1$   & $1.1462$ & $2.35$ \\ 
  \end{tabular}
  \caption{The roughness exponent for depinning $\zeta_{dep}$ at the point of minimum sensitivity upon varying the parameter $a$ in the dimensionless IR cutoff function $\hat r(\hat q^2)=(a+\frac{\hat q^2}{2})e^{-\hat q^2}$.}\label{tab1}
\end{center}
\end{table}

We have chosen an IR cutoff function of the (dimensionless) form  $\hat r(\hat q^2)=(a+\frac{\hat q^2}{2})e^{-\hat q^2}$ with one free parameter $a$. The sensitivity of the results to the choice of IR cutoff function can be estimated by studying their dependence on the parameter $a$. For illustration, we give the ``optimal'' value $a_{PMS}$, which is determined here by the minimum in the dependence $\zeta(a)$, in the case of depinning for several dimensions: see Table~\ref{tab1}. One can see that $a_{PMS}$ varies with $d$ (it also changes between equilibrium and depinning). However, the variation of the exponents and fixed-point functions with $a$ is very slow. For instance, if we fix $a=a_{PMS,d=3}=4.8$ for all dimensions $d$, we find that the change in $\zeta$ compared to the result at the point of minimum sensitivity with $a_{PMS,d}$ is less than $0.5 \%$ in $d=1$ (with $\zeta=1.151$ instead of $1.146$), less than $0.1\%$ in $d=2$ (with $\zeta=0.7587$ instead of $0.7581$), and is not even visible at the fifth digit in $d=3.9$. This robustness of the results for a whole range of parameters parametrizing the IR cutoff function is in line with what is generically found in NP-FRG studies, in particular for the case of the RFIM.\cite{balog_eqnoneq}

\section{Multicritical fixed points for the random-bond disorder in equilibrium}
\label{app_multicriticalFP}

In this appendix we provide some details concerning the search for fixed points and the study of their stability in the equilibrium case for a random-bond disorder.

Consider first the dimensionless NP-FRG equation for the second cumulant of the random force in the equilibrium case, which we have derived in Sec. \ref{sec_NPFRG}:
\begin{equation}
\label{del_simple}
\partial_t\delta_k(y)=(d-4+2\zeta)\delta_k(y)-x\zeta\delta_k'(y) +\frac{1}{2}I_3\big([\delta_k(y)-\delta_k(0)]^2\big)'' +\frac{\sqrt{2}}{2}I_2 \partial_y w_k(0,\frac{2y}{\sqrt{3}})\,,
\end{equation} 
where we have used the symmetry characteristic of equilibrium which implies that $w_k(y,y/\sqrt{3})=-w_k(0,2y/\sqrt{3})$. As stressed in the main text, integrating both sides of the equation over $y$ between $0$ and $+\infty$ leads, with minor assumptions about the behavior of the cumulants at large $y$, to the property that $\partial_s \int_0^{+\infty}dy\,\delta_k(y)=(d-4+3\zeta) \int_0^{+\infty}dy\,\delta_k(y) $. For random-bond disorder  $\int_0^{+\infty}dy\,\delta_{k}(y)=0$ at the beginning of the flow (when $k=\Lambda$) and it therefore stays equal to zero all the way to the fixed point.

Another feature at equilibrium, which also derives from the fact that the renormalized random force derives from a renormalized random potential and from the original $Z_2$ symmetry, is that $w_k(0,0)=0$. It is easily checked that this property is conserved along the NP-FRG flow equation. The physical fixed points in the random-bond universality class(es) must therefore satisfy at least the two above properties.

\begin{figure}[h!]
   \begin{center}
     \includegraphics[width=300pt,height=240pt]{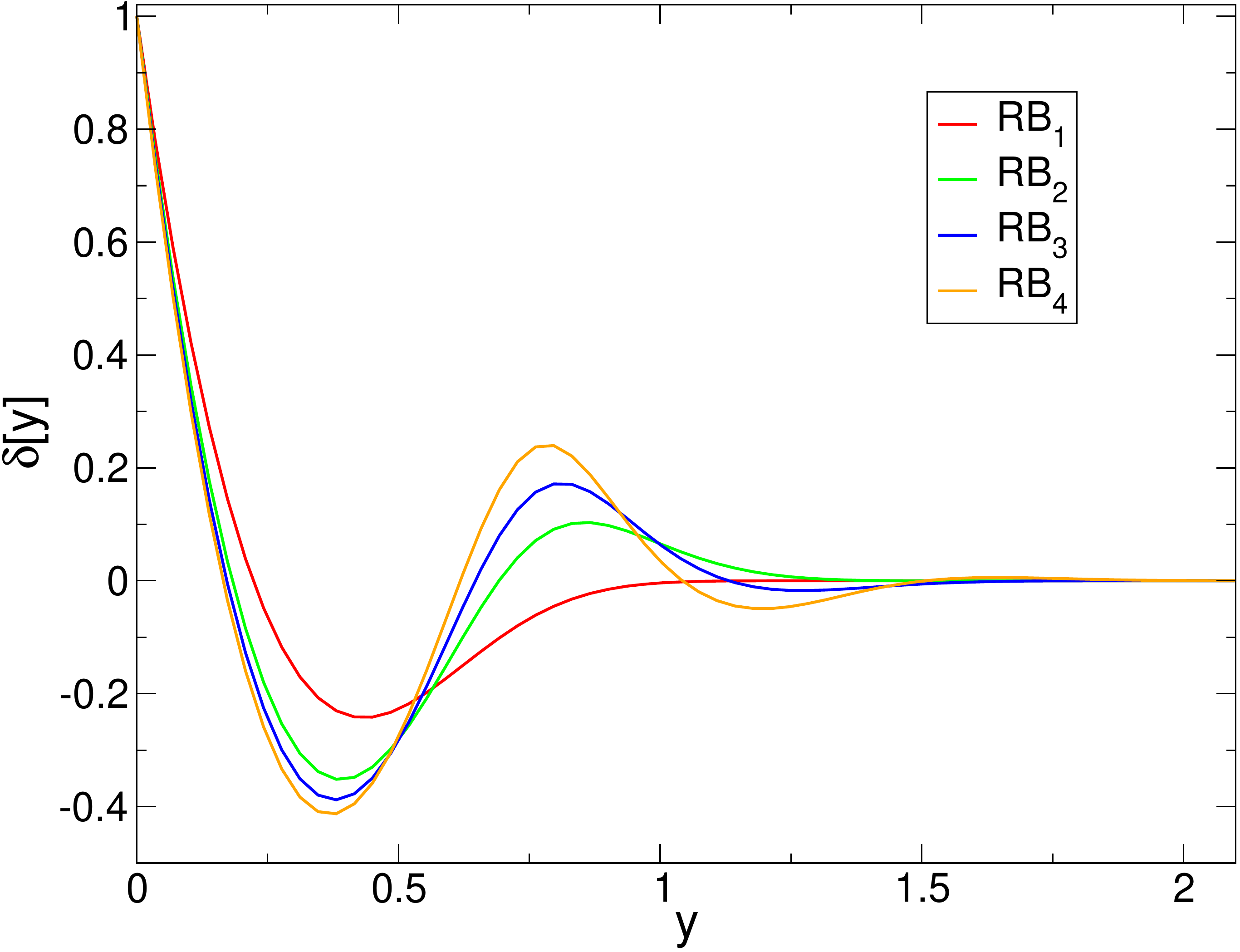}  
     \caption{The fixed-point solutions for the function $\delta_*(y)$ for a pinned manifold in equilibrium in a random-bond disorder for $d=3$. $RB_p$ denotes the fixed point with $p$ nodes.}
   \label{fig_RBFPS_d3}
    \end{center}
   \end{figure}

By numerically solving the coupled fixed-point equations for $\delta_*(y)$ and $w_*(y,z)$ for $0\leq d \leq 4$, as discussed in the preceding appendix, we have found several random-bond fixed points. They all satisfy $\int_0^{+\infty}dy\,\delta_{*}(y)=0$ and $w_*(0,0)=0$. For illustration we display the fixed-point functions $\delta_{*}(y)$ for $d=3$ in Fig. \ref{fig_RBFPS_d3}. As can be seen the function $\delta_{*}(y)$ has oscillations  and one find solutions with an increasing number of nodes. The associated roughness exponents are found to be $\zeta=0.21$, $0.14$, $0.11$, and $0.09$ for the fixed points with $1$, $2$, $3$, and $4$ nodes, respectively. (There are solutions with more nodes but accessing them requires the use of larger grids for the numerical resolution, which, knowing that these fixed points have an increasing number of unstable directions, we did not find worth studying.)

\begin{figure}[h!]
   \begin{center}
     \includegraphics[width=300pt,height=240pt]{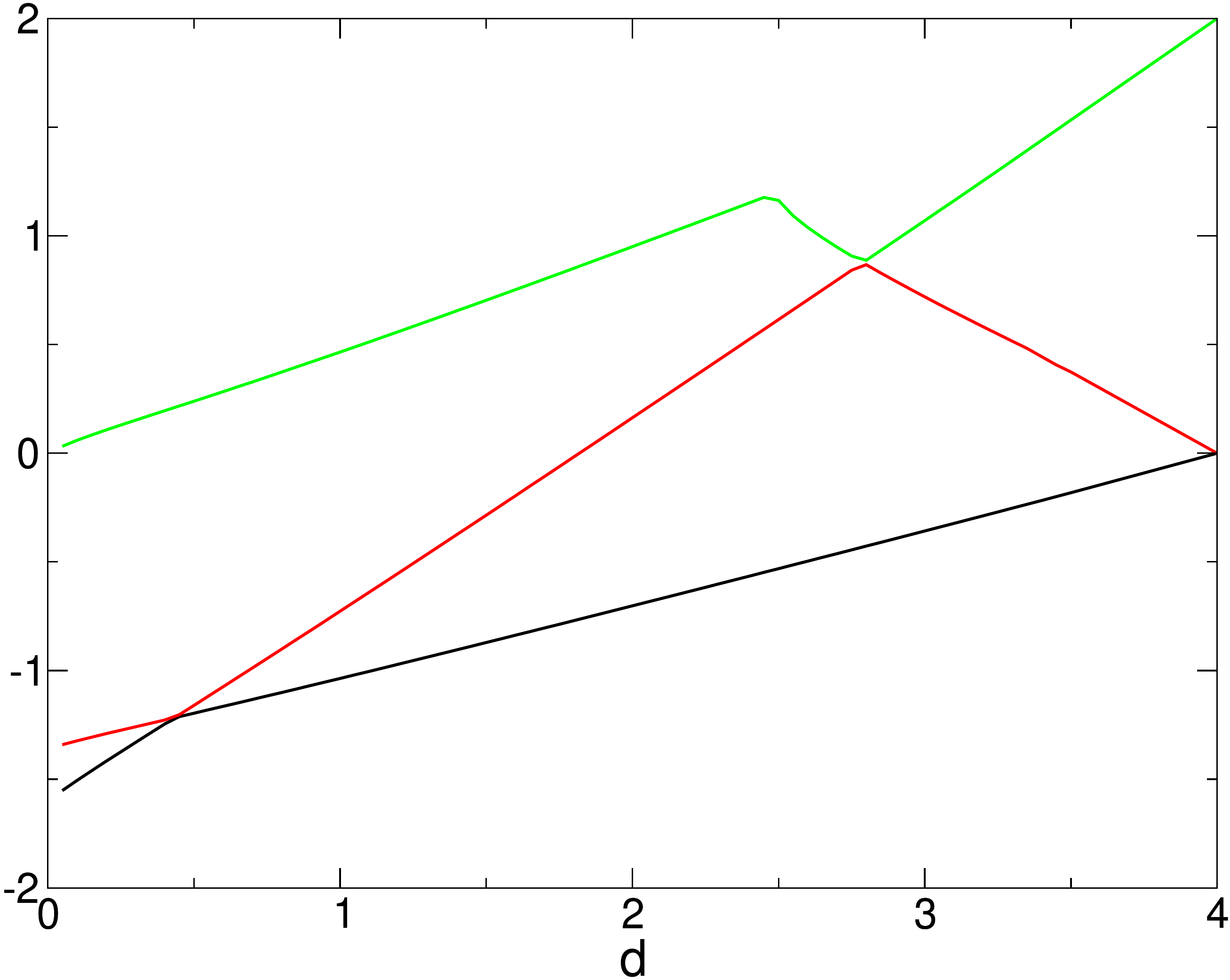}  
     \caption{The lowest 3 eigenvalues of the stability matrix around the random-bond fixed point with 1 node in $\delta_*(y)$. The relevant (negative) eigenvalues correspond to unphysical directions. The fixed point is therefore fully attractive in the physical subspace.}
     \label{fig_RB1_stability}
   \end{center}
 \end{figure}

To better characterize these random-bond fixed points we have studied their stability. Consider first the $1$-node fixed point. We plot in Fig. \ref{fig_RB1_stability} the lowest eigenvalues obtained by diagonalizing the stability matrix derived from the coupled NP-FRG flow equations around the fixed point. For dimensions greater than $\approx 2$ it has only a single unstable direction which breaks potentiality ({\it i.e.}, it is such that $\int_0^{+\infty}dy\,\delta_k(y) \neq 0$) and is thus unphysical. Below $d \approx 2$, there appears another unstable direction. This perturbation only affects the third cumulant and is such that $w_k(0,0)\neq 0$, so that it is also unphysical. In the physical subspace, the fixed point with 1 node in $\delta_*(y)$ is therefore fully attractive. This is the stable fixed point found by the perturbative FRG, which we discuss in the main text. 

We find that the number of unstable directions increases with the number of nodes characterizing $\delta_*(y)$ at the fixed point. Excluding the directions that violate the physical requirements that $ \int_0^{+\infty}dy\,\delta_k(y)=w_k(0,0)=0$, we obtain that the fixed point with 2 nodes is once unstable, that with 3 nodes is twice unstable, etc. The physical meaning of these critical and multi-critical fixed points which seems to be present in all dimensions is unclear to us. As far as we know there is no reported observation of such critical features in the equilibrium behavior of a manifold pinned by a random-bond disorder in either experimental or simulation studies.

\section{Expression for the third cumulant $w_k(y,z)$ at leading order when $k\to 0$ and at lowest order in $\epsilon=4-d$}
\label{app-epsilon_exp}

By solving the flow equation for the third cumulant at leading order near the fixed point and at the lowest order in $\epsilon$, we obtain the following expression (for $y\geq 0$ and $z\geq y/\sqrt 3$):
\begin{equation}
\begin{aligned}
\label{w_epsilon3}
&w_k(y,z)\approx\frac{3\sqrt{2}}{8} I_4 \Big\{-\delta_k'(y)^2 \big [\delta_k'(\frac{1}{2} (-y+\sqrt{3} z))+\delta_k'(\frac{1}{2} (y+\sqrt{3} z))\big ] \\
&+\delta_k '(\frac{1}{2} (-y+\sqrt{3} z))^2 \delta_k '(\frac{1}{2} (y+\sqrt{3} z))+\delta_k '(y) \Big (\delta_k ''(\frac{1}{2} (-y+\sqrt{3} z))\big [-\delta_k (0)+\delta_k (y) \\ 
&+\delta_k (\frac{1}{2} (-y+\sqrt{3} z))-\delta_k (\frac{1}{2} (y+\sqrt{3} z))\big ]+ \delta_k ''(\frac{1}{2} (y+\sqrt{3} z)) \big [\delta_k (0)-\delta_k (y) \\
& +\delta_k (\frac{1}{2} (-y+\sqrt{3} z))-\delta_k (\frac{1}{2} (y+\sqrt{3} z))\big ] 
+\big [\delta_k '(\frac{1}{2} (-y+\sqrt{3} z))^2-\delta_k '(\frac{1}{2} (y+\sqrt{3} z))^2\big ] \Big )\\
&+\delta_k '(\frac{1}{2} (-y+\sqrt{3} z)) \Big (\delta_k ''(y) \big [\delta_k (0)-\delta_k (y) -\delta_k (\frac{1}{2} (-y+\sqrt{3} z))+\delta_k (\frac{1}{2} (y+\sqrt{3} z))\big ]
\\&+ \delta_k ''(\frac{1}{2} (y+\sqrt{3} z))\big [-\delta_k (0)-\delta_k (y)+\delta_k (\frac{1}{2} (-y+\sqrt{3} z)) +\delta_k (\frac{1}{2} (y+\sqrt{3} z))\big ] +\delta_k '(\frac{1}{2} (y+\sqrt{3} z))^2 \Big )
\\&+\delta_k '(\frac{1}{2} (y+\sqrt{3} z)) \Big (\delta_k ''(y) \big[\delta_k (0)-\delta_k (y)+\delta_k (\frac{1}{2} (-y+\sqrt{3} z))-\delta_k (\frac{1}{2} (y+\sqrt{3} z))\big ]
\\& +  \delta_k ''(\frac{1}{2} (-y+\sqrt{3} z)) \big [-\delta_k (0)-\delta_k (y)+\delta_k (\frac{1}{2} (-y+\sqrt{3} z))+\delta_k (\frac{1}{2} (y+\sqrt{3} z))\big ]\Big )
\\&+ 2\lambda \delta_k'(0^+)\Big (\delta_k'(y)\big [\delta_k '(\frac{1}{2} (\sqrt{3} z+y))-\delta_k '(\frac{1}{2} (-y+\sqrt{3} z)) \big ] +\delta_k '(\frac{1}{2} (y+\sqrt{3} z))\delta_k '(\frac{1}{2} (-y+\sqrt{3} z))\Big )\Big\}
\\& +{\rm O}(\epsilon^4)\,,
\end{aligned}
\end{equation}
where near the fixed point $\delta_k(y)\sim \epsilon$ so that $w_k(y,z)\sim \epsilon^3$. Recall that $\lambda=0$ for equilibrium and $\lambda=1$ for the depinning case. The above scaling solution for the third cumulant can now be inserted in the flow equation for the second cumulant.


\begin{thebibliography}{99}
\bibitem{footnote_perturbRFIM}
Keeping track of the avalanches has been shown to be possible in a  perturbative FRG of the RFIM in $\epsilon=6-d$.\cite{tissier_perturb} However, a detailed knowledge of the behavior at and below the critical dimension $d_{DR}$ requires a nonperturbative approach.
\bibitem{footnote_REMM}
One should not confuse the acronym REMM introduced here with a very different glassy disordered model, the random energy model, known as REM.
\end{thebibliography}
\end{document}